# Scilit with the Integrated Impact Indicator Assessment

**Haochen Dong[1,*], Sun Qiao[1,*], Yanping Mu[1], Lu Liao[1], Diogo Rodrigues [2], Frank Sauerburger [2], Yi Bu[3], Robin Haunschild[3,4]**

**1** MDPI, Poly Metropolitan Floor 10-14, Building 2, Courtyard 4, Guanyinan North Street, Beijing 101101, China;
2 MDPI, Grosspeteranlage 5, 4052 Basel, Switzerland;
**3** Department of Information Management, Peking University, 5 Yiheyuan Road, Haidian District, Beijing 100871, China;
4 Max Planck Institute for Solid State Research, Heisenbergstr. 1, 70569 Stuttgart, Germany
**Correspondence:** MDPI, Poly Metropolitan Floor 10-14, Building 2, Courtyard 4, Guanyinan North Street, Beijing 101101, China. dhcydll1004@gmail.com; stevensunss@163.com

Abstract

In this study, we systematically elucidate the background and functionality of the Scilit database and evaluate the feasibility and advantages of the comprehensive impact metrics I3 and I3/N, introduced within the Scilit framework. Using a matched dataset of 17,816 journals, we conduct a comparative analysis of Scilit I3/N, Journal Impact Factor, and CiteScore for 2023 and 2024, covering descriptive statistics and distributional characteristics from both disciplinary and publisher perspectives. The comparison reveals that the Scilit I3 and I3/N framework significantly outperforms traditional mean-based metrics in terms of coverage, methodological robustness, and disciplinary fairness. It provides a more accurate, diagnosable, and responsible solution for interdisciplinary journal impact assessment. Our research serves as a “getting started guide” for Scilit, offering scholars, librarians, and academic publishers in the fields of bibliometrics or scientometrics a valuable perspective for exploring I3 and I3/N within an inclusive database. This enables a more accurate and comprehensive understanding of disciplinary development and scientific progress. We advocate for piloting and validating this method in broader evaluation contexts to foster a more precise and diverse representation of scientific progress.

## 1. Introduction

In the current research ecosystem, scholarly communication, scientific efforts, and academic reputation are closely linked to researchers' productivity. As the main channels of academic output, journals gain from impact evaluations based on bibliometric metrics provided by indexing databases like Web of Science (WoS) (Birkle, Pendlebury, Schnell, & Adams, 2020) and Scopus (Baas, Schotten, Plume, Côté, & Karimi, 2020). As representatives, Journal Impact Factor (JIF) from WoS (Clarivate) (Clarivate, 2025b; Zhou & Leydesdorff, 2011) a two-year average number of citations in year Y to the total number of citable items from years Y−1 to Y−2, while CiteScore (Feldner, 2025; Meho, 2019) from Scopus (Elsevier) uses a four-year window (i.e., from Y−1 to Y−4) with broader document coverage and rolling monthly updates, culminating in an annual snapshot; both are non-field-normalized citation indicators best suited for within-field journal benchmarking.

Evaluations of journals' citation impact have led to the widespread use of citation-based indicators across the scholarly ecosystem (Wilsdon, 2016). Most directly, such indicators support journal stratification and disciplinary rankings (Ahmad, et al., 2020), thereby serving as a key reference for researchers when choosing target journals for submission. From the perspective of publishers, prominently featuring citation metrics has become a central strategy for brand communication and market positioning, aimed at attracting high-quality submissions and readership. For academic libraries, citation indicators likewise constitute a core reference in journal selection and renewal decisions. In addition, within universities and research institutes, these indicators are frequently employed as proxy measures of scholarly competence in hiring, promotion, and tenure evaluations; at national or institutional levels, they are also used as reference criteria in the allocation of performance bonuses and research funding. Nevertheless, when applied to the assessment of individual researchers or to resource distribution, such practices risk the inappropriate extrapolation of journal-level metrics to the level of persons and projects, a use that remains contested. Despite sustained criticism, citation-based indicators continue to dominate evaluative practices due to their perceived simplicity, scalability, and quantitative comparability; however, structural issues such as database coverage biases, field-dependent citation densities, highly skewed citation distributions, numerator–denominator asymmetries, and the strategic behavior of stakeholders (for example, editorial policies and citation gaming) underscore the need to adopt more scientifically grounded citation indicators (Fong & Wilhite, 2017; Juyal, Thawani, Sayana, & Pal, 2019; Wilhite & Fong, 2012).

The old bulls WoS and Scopus have been confronted with competitors. Joshi (2016) provides a structured overview of Scopus and WoS features and coverage, emphasizing Scopus' breadth across journals, books, conferences, and patents, and WoS' established citation-linking and analytics. Similarly, Aghaei Chadegani, et al. (2013) synthesize comparative literature on Scopus and WoS and find no universal winner regarding the choices of their field, time span, and use case (search vs evaluation). Based on their observations, Scopus often covers more journals, while WoS frequently leads in historical depth and some analytics. In particular, Google Scholar (https://scholar.google.de/) (Martín-Martín, Orduna-Malea, Thelwall, & López-Cózar, 2018), Dimensions (https://www.dimensions.ai/) (Singh, Singh, Karmakar, Leta, & Mayr, 2021), and OpenAlex (https://openalex.org/) (Priem, Piwowar, & Orr, 2022) have emerged as free alternatives. However, Google Scholar is more a search engine than a bibliographic citation database and can't be classified as an open research information data source (ORIDS). Adriaanse and Rensleigh (2013) studied South African environmental science journals (2004–2008) and found that WoS retrieved the highest total citations and the most unique items; Scopus delivered the best metadata consistency, while Google Scholar exhibited the most issues and multiple copies. They detected no duplicate records in WoS or Scopus, whereas Google

Scholar showed many duplicates and inconsistencies in author order, volume/issue, and pagination. Access to Dimensions is not completely free. Some features require a paid subscription. Downloads are restricted. Thus, Dimensions also doesn't count as a true ORIDS. However, Google Scholar and Dimensions are more inclusive in their coverage than WoS and Scopus. OpenAlex is even more inclusive with regard to scientific publications than Dimensions and is a true ORIDS. The coverage of Google Scholar is probably larger than the one of OpenAlex but this is only assessable for test cases as Google Scholar is not available for download as a snapshot. However, the meta data quality of Google Scholar is questionable as well as the scientificity of the indexed documents.  concluded that search engines such as Google Scholar have a lower meta data quality than bibliographic citation databases such as OpenAlex. Thus far, OpenAlex mainly relies on donations and grants besides fees for premium services in order to be maintained sustainably.

Despite the growing availability of bibliometric databases, existing platforms face several limitations. Comprehensive profiles for key academic actors—scholars, institutions, and publishers, which are often incomplete, and detailed information about their outputs and relationships is limited. Besides, lots of databases classify publications only in general categories, without distinguishing more specific subtypes. Moreover, author metadata is frequently ambiguous due to limited disambiguation, which hinders accurate attribution of publications and metrics. Access to data is often restricted with a cost, further limiting usability.

Indeed, the scope of the database's inclusion and meta data quality influences the calculation of the metrics. Non-exclusionary databases, on the other hand, contain a larger and more comprehensive network of citations, allowing for a more refined understanding of the impact of citations at the journal level. Scilit, developed by Multidisciplinary Digital Publishing Institute (MDPI), aggregates scholarly records from a broad array of sources to maximize coverage and timeliness. Its ingestion pipeline primarily harvests metadata from DOI registration agencies (notably Crossref and DataCite), integrates records from PubMed/Medline for biomedical literature, and indexes content supplied directly by publishers via feeds and APIs. In addition, Scilit incorporates open-access directories (e.g., DOAJ) and selected preprint servers, while routinely synchronizing publisher corrections and updates to maintain currency. This multi-source, non-exclusive model, unconstrained by a proprietary, closed indexing policy, enables Scilit to include journals from large and small publishers, society titles, regional and non-English outlets, conference proceedings, book series, and preprints. In other words, Scilit addresses these gaps by aggregating data across multiple sources to provide more complete profiles of scholars, institutions, sources, and publishers. It applies advanced author disambiguation techniques to resolve ambiguities and classifies publications into 14 specific subtypes, including Research Article, Review Article, Conference Paper, Clinical Trial, Case Report, Letter, Book Review Abstracts, Editorials, Retracted Article, Retraction, Correction, Withdrawal,  and Other, which enables users to perform refined searches and retrieve detailed, reliable information.  Furthermore, Scilit is fully free to use, and its operation by MDPI ensures financial support and sustainable development, providing broad and reliable access to comprehensive academic data. As a result, its coverage is intentionally broader than many selective or subscription-based databases, improving discoverability across disciplines, languages, and publication formats and reducing biases associated with narrow curation criteria. Currently, the database contains more than 181 million entries, mainly journal papers, conference proceedings papers, books, book-chapters, and preprints, and a data snapshot for Scilit and OpenAlex is presented in Table 1. Overall, OpenAlex includes more works than Scilit. However, not all works in OpenAlex are scholarly material. One known shortcoming of OpenAlex is incomplete reference matching. The higher numbers of citations per paper and citing papers per paper of Scilit than in OpenAlex show that Scilit has a higher completeness in reference matching than OpenAlex. Accurate and complete reference matching is very important for the calculation of citation-based indicators.

Table 1. Comparative overview of Scilit and OpenAlex data across multiple dimensions within 2015-2024.

| | **OpenAlex (2015-2024)** | **Scilit (2015-2024)** |
|---|---|---|
| Count of total works | 100,059,695 | 72,244,250 |
| Count of works with author or institution: | 91,405,199 | 39,348,564 |
| Count of authors with at least one paper | 53,186,532 | 15,702,524 |
| Count of institutions with at least one paper | 93,606 | 73,376 |
| Count of works with at least one author | 90,134,941 | 31,519,239 |
| Share of works with at least one author | 0.901 | 0.436 |
| Count of works with at least one institution | 49,430,281 | 21,620,963 |
| Share of works with at least one institution | 0.494 | 0.299 |
| Count of works with at least one author AND at least one institution | 49,427,038 | 30,513,489 |
| Share of works with at least one author AND at least one institution | 0.494 | 0.422 |
| Count of citation relations | 549,809,112 | 1,529,927,725 |
| Citation relations per paper | 5.495 | 21.177 |
| Count of cited papers | 36,332,876 | 30,765,138 |
| Cited papers per paper | 0.363 | 0.426 |
| Count of citing papers | 36,702,097 | 31,167,211 |
| Citing papers per paper | 0.367 | 0.431 |
| article | 66,707,420 | 46,056,219 |
| dataset | 5,763,661 | 1,542,655 |
| preprint | 4,471,388 | 4,226,705 |
| books | 1,328,610 | 1,005,810 |
| book-chapters | 10,694,341 | 11,529,965 |
| editorial | 305,838 | 1,034,235 |
| retraction | 19,577 | 38,370 |
| thesis | 2,710,536 | 524,211 |
| reference-entry | 288,908 | 181,897 |
| retracted-article | 31,925 | 41,522 |

On the other hand, Integrated Impact Indicator (I3) (Leydesdorff & Bornmann, 2011) offers a percentile-based, distribution-sensitive alternative to JIF and CiteScore for evaluating journal influence. By aggregating article-level citation performance across the full citation distribution using percentile ranks, I3 mitigates distortions arising from skewed citation distributions, avoids numerator–denominator asymmetries inherent in averages, and enables field-normalized comparisons through reference sets. Computed as the sum of weighted publication counts across citation-percentile classes, I3 captures both volume and impact without being dominated by a few highly cited outliers, thereby providing a more statistically robust and scientifically grounded measure of journal impact suitable for cross-field assessment and responsible research evaluation.

Although existing studies have systematically compared the performance of JIF and CiteScore within WoS/Scopus, there remains a lack of systematic evidence regarding the applicability and advantages of the distribution-sensitive, field-normalized I3 metric in non-exclusive, broad-coverage databases. Particularly for fields with high proportions of non-English journals, regional journals, conference proceedings, and book chapters, existing mean-based metrics may amplify biases stemming from differences in database coverage and disciplinary citation density. A new empirical evaluation framework is urgently needed. Building on the foregoing context, this study pursues two objectives: i) to introduce the Scilit database and its core functionalities; and ii) to implement and compute I3 within the Scilit database, benchmarking it against JIF and CiteScore to assess its suitability as a journal impact metric. The contribution of this work is twofold. First, it is the first to integrate Scilit with the I3 framework. Second, by calculating and comparing indicators (i.e., JIF and CiteScore) for the years 2023 and 2024, it demonstrates the advantages of deploying a percentile-based, distribution-sensitive metric within a non-exclusionary, multi-source database. Together, these results highlight the methodological fit between I3 and Scilit's expansive coverage and provide an alternative pathway for responsible, cross-field journal impact evaluation.

## 2. Introduction of Scilit database

### *2.1 Scilit History*

Scilit, was proposed as "OAlit" in the fall of 2012 and developed by MDPI to support the open-access publishing ecosystem, the goal was to collect all open-access papers (including the full-text) and build a searchable open database. The scope was later extended to all scientific papers in 2014, and the service was renamed to Scilit, it was scraping metadata from CrossRef API, PubMed, PMC, Preprint server platforms including arXiv, HAL, PeerJ, and RePEc.

The service has been continuously improved and expanded by scraping data from an increasing number of publisher websites, collaborating with publishers who deposit metadata into Scilit, integrating additional data sources, enhancing scraping algorithms, and improving metadata disambiguation. Following the introduction of Crossref citation data in 2014 and its enhancement in 2022, Scilit progressively extended its citation dataset by integrating citation and reference data.

### *2.2 Data Sources and Architecture*

Data ingestion is mainly performed from authoritative sources like Crossref, PubMed, PMC, and DOAJ (Fig. 1). Harvesting of metadata as a real-time mechanism is proceeded with aggregation, deduplication, and disambiguation (especially authors and institutions) to deliver high-quality data.

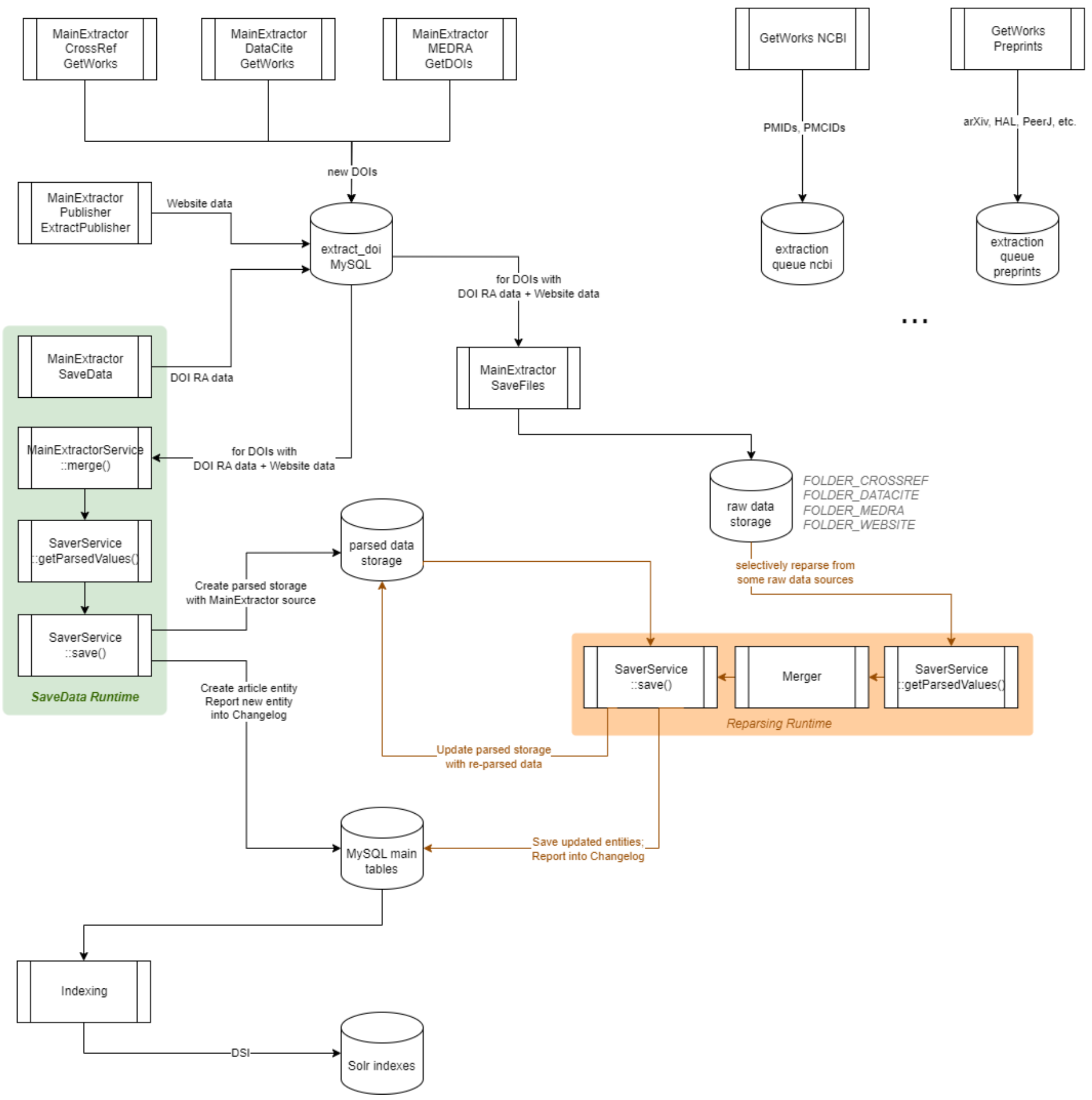

Figure 1. Scilit data processing diagram

*2.3 Search Functionalities*

Scilit offers both basic and advanced search functionality. The basic Scilit search provides a fast way to discover scholarly research content. Users can simply enter keywords, the system automatically searches in the title, keywords, abstract, DOI, and ROR ID fields of all indexed publications, Scilit will return the relevant research, users can also refine the results based on the multiple filter facets, like publication type, categorizations of subject disciplines, SDG classifications, publication year, country/region, publisher, source, and MeSH headings to retrieve more targeted results. The advanced search function in Scilit enables refined querying across a wide range of metadata fields, including title, abstract, keywords, subject, DOI, ROR ID, author name, source title, publisher, and publication year. It supports complex and structured literature exploration through flexible field selection. In addition, the advanced search panel allows the use of Boolean operators: AND, OR, and NOT — helping users to construct specific and precise queries.

*2.4 Scilit Export Options*

Scilit supports users to export search results in the format of CSV, RIS, EndNote, BibTeX, and a publication list file. Exported results can include the following fields: Publication Title, Authors, Source Title, Publisher, Publication Keywords, Volume, Page, Open Access status, Times Cited, Article DOI, Date Published, and Scilit Article Page Link. Exports are limited to 1000 records at a time but can be paged. Access via an application programming interface and a downloadable snapshot are planned.

*2.5 Scilit Rankings*

Scilit provides a comprehensive ranking system that highlights the performance of various entities in scholarly publishing. The rankings allow users to 1) track the publication volume and impact of top-performing entities; 2) compare entities across output and impact; 3) identify emerging journals, active organizations, or influential preprint platforms.

In addition, Scilit offers a comparison tool, enables users to select multiple entities and view side-by-side metrics, trends, and ranking changes over time. This feature is especially useful for research assessment, institutional benchmarking, and editorial strategy. It includes the following highlights:

1. Scilit rankings cover 5 kinds of dimensions, including Publishers, Journals, Preprint Servers, Organizations, and Countries. Each sub-ranking section is calculated mainly based on publishing output and citation metrics as follows:
   i) Publishing Output – the total number of publications/OA publications in one edition year (Edition year of Scilit Rankings and equals the last full year of data we have on file, i.e. as of July 2021 the EDY is 2020.) in Scilit;
   ii) Citation Metrics – including 5-Year Cited Articles, 5-Year Citations, h5-index, Monthly Citation Metrics.
2. Scilit Rankings provide a functionality of comparing: In this part, users can conduct a comparison directly based on publishing performance among up to three entities of publishers, journals, and organizations.
3. Publishing market: Users can also have an overall view on the academic publishing market developments regarding the number of articles, journals, and preprints, as well as having an insight on the contribution percentage by publishing houses.

*2.6 Scilit Subject Classification*

Scilit subjects, developed by MDPI, ensures a comprehensive and robust two-level categorization system covering all disciplines published by the organization. The 12 high-level categories are: Agriculture; Arts & Humanities; Biology & Life Sciences; Chemistry & Materials Science; Clinical Sciences; Computer & Information Science; Earth & Environmental Sciences; Electrical Engineering & Electronics; Engineering; Mathematics; Physics; Social Sciences. Each high-level category is further divided to create a total of 290 subcategories that refine the scope, such as Catalysts, Nanoparticles, or Textile Chemistry within Chemistry & Materials Science. This hierarchical structure enables precise classification of interdisciplinary research while maintaining a clear overview of each academic domain.

To assign these subjects, we employ a Natural Language Processing (NLP) classification model based on an in-house fine-tuned BERT model called ScilitBERT (de la Broise, Bernard, Dubuc, Perlato, & Latard, 2021). ScilitBERT is pre-trained on a diverse and extensive corpus of academic text across multiple disciplines, and its architecture includes a classification layer added to the base network for subject assignment. The model uses articles' title and abstract to infer the subject(s) of each articles' content.

This AI-powered model functions as a multi-label classifier, allowing manuscripts to be assigned to one or multiple categories. The subject categories of a journal are built by aggregating the categories of their articles from the past two years, ensuring they reflect the journals' real content. This approach provides an objective, data-driven representation of the journals' scope.

*2.7 The entities and ER diagrams produced by Scilit*

Scilit is organized in a relational database structure, with publications as the central entity linked to Scholars, affiliations, sources, publishers, and references. Entity–Relationship (ER) diagrams are provided in Figure 2 - Figure 7, illustrating the key entities and their relationships.

*2.7.1 Scholar Entity*

The author system is designed to establish academic archives for scholars, including their basic information, academic achievements, academic relationships, research fields, published papers, and related indicators, to showcase their research achievements and academic contributions.

There are three primary entities UniqueAuthor, UniqueAuthorExtraInfo, LogUniqueAuthorDisambiguation. UniqueAuthor is the central entity for the author system. It is generated by the Authors entity and aggregates core author details like name, email, ORCID, affiliations, and institutions. Some attributes like names, emails, affiliations, and institutions are split into separate entities, to optimize storage and reusability. For example, AuthorEmails manages authors' email addresses, Affiliations capture specific author-institution relationships, Institutions store detailed information about research institutions, including name, country, and website. UniqueAuthorExtraInfo tracks academic metrics such as publications, citations, self-citations, h-index, i10-index, academic age, subjects, and its ranking （ Subjects and UniqueAuthorSubjects define research domains, and UniqueAuthorSubjectsRanking records authors' rankings within their research fields). LogUniqueAuthorDisambiguation monitors and records the authors of identity conflicts and the progress of disambiguation, including email, status (disambiguation progress), and note (additional context).

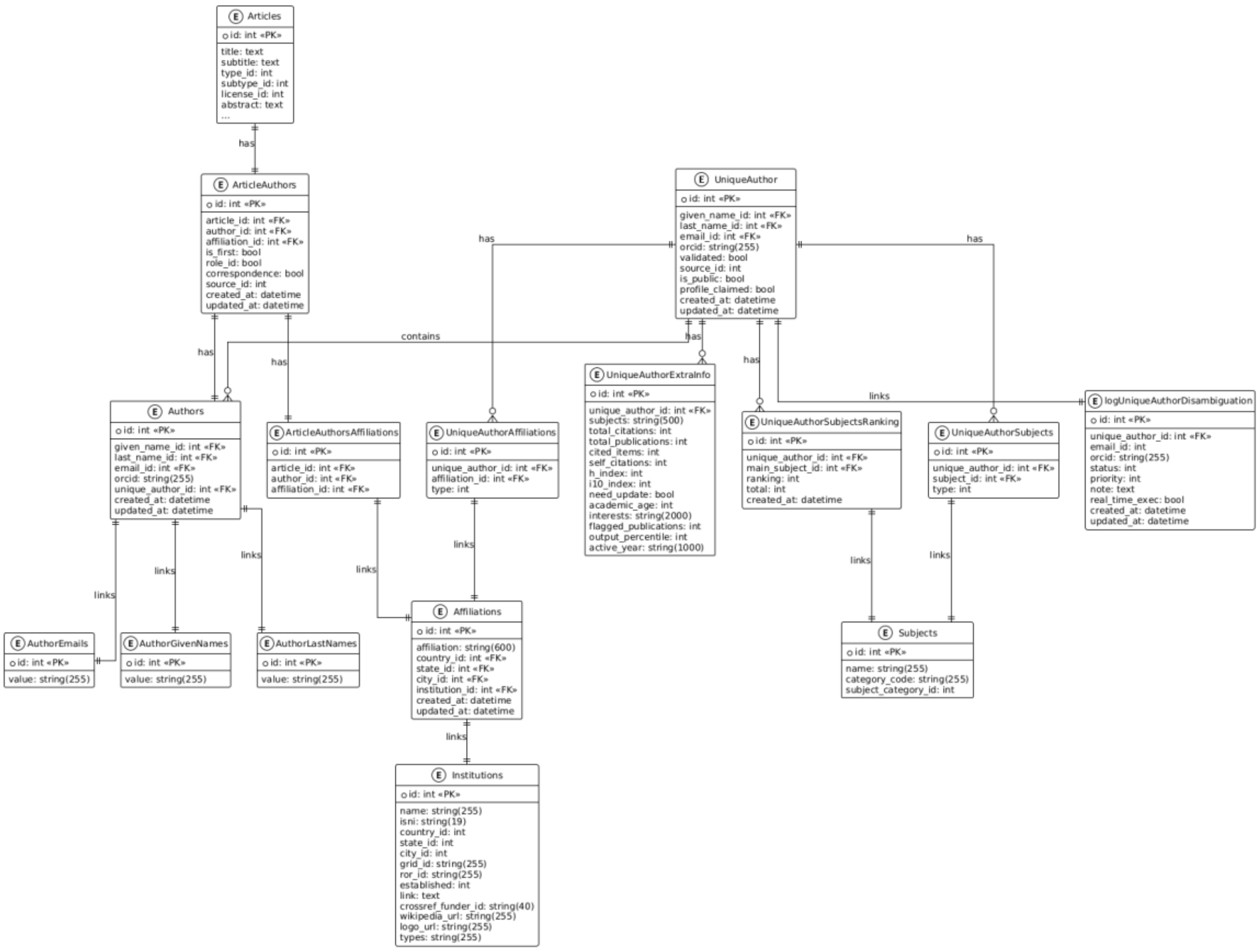

Figure 2. ER diagram of the Scholar Entity

To ensure high-quality scholar metadata, Scilit implements a two-stage validation for author information: an Author Disambiguation Model and an Ambiguity Detection Model.
The author disambiguation model focuses on merging records that belong to the same individual, while the ambiguity detection model identifies records that may require further human or algorithmic clarification due to insufficient metadata. Together, these models form the foundation for reliable author profiles, accurate affiliation links, and stable scholar metrics.

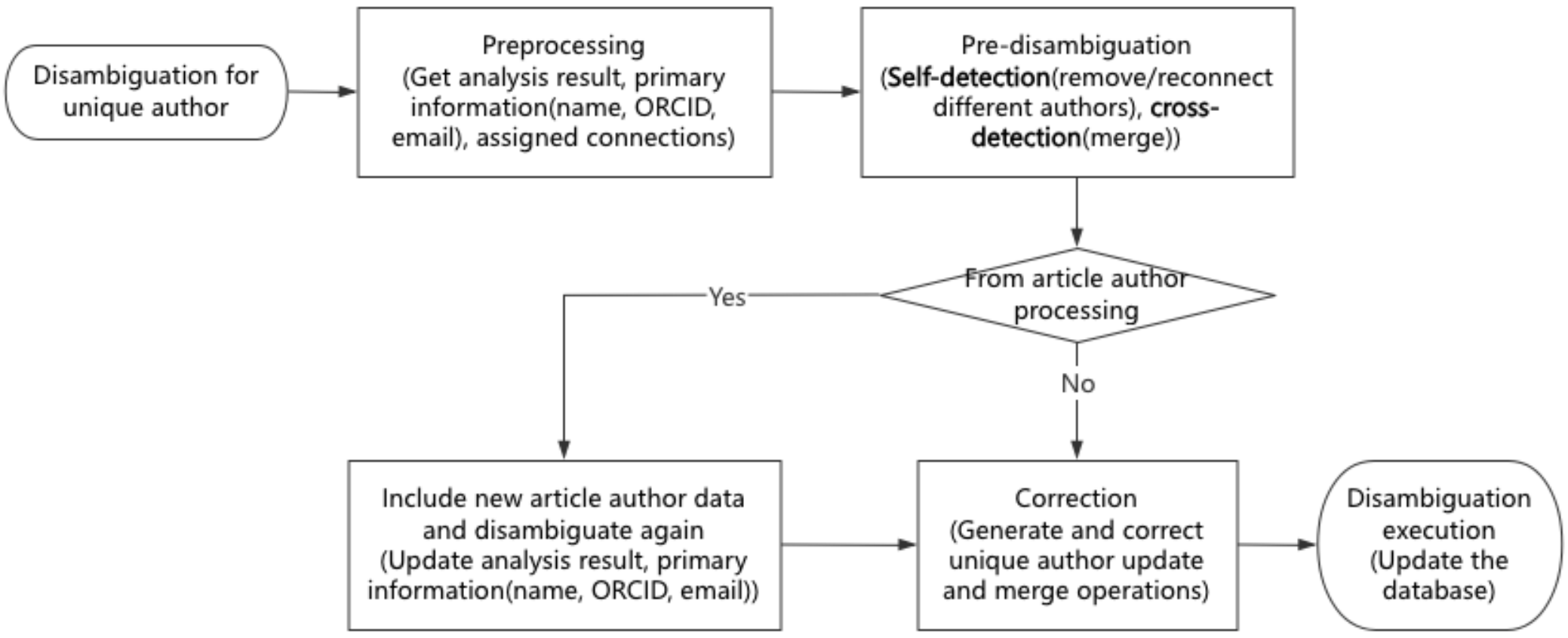

Figure 3. Author Disambiguation Model.

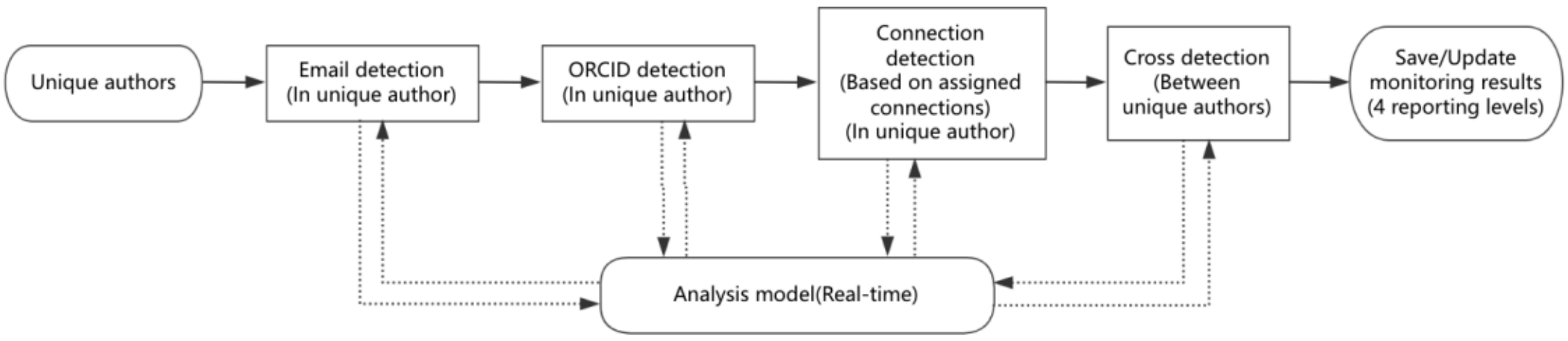

Figure 4. Ambiguity Detection model

Building upon our existing disambiguation and ambiguity detection models, the AI team is currently developing and applying more advanced approaches to further enhance author metadata. Leveraging the existing dataset as a foundation, these methods are expected to substantially improve the accuracy, completeness, and reliability of author information, thereby providing a stronger basis for downstream analyses and metrics.

### *2.7.2 Sources Entity*

Source entity typically indicated journal, preprint, proceedings, books, or book-chapters. The system is a sophisticated content aggregation platform designed to manage and organize scholarly publications into logical containers. The system is built around two primary entities: WorkContainer and WorkContainerGroup, where individual containers (like specific journal issues or preprint collections) are grouped under broader container groups (like a journal series or conference series). The architecture follows the Factory Method design pattern to handle different container types (journal, preprint, proceedings, and books) with a specialized processing logic for each type. Each container entity includes metadata such as titles, abbreviations, ISSNs, publication types, and relationships to publisher groups and container groups, while the service layer (WorkContainerService) coordinates the processing, validation, and persistence of container data, including conflict resolution when multiple sources provide conflicting information about the same container. The system also incorporates Solr-based search capabilities, conflict detection mechanisms, and post-processing workflows to maintain data quality and ensure proper indexing of scholarly content. Additionally, it includes observer pattern implementations for event handling, citation metrics tracking, and comprehensive repository methods for efficient data access and management across the container ecosystem.

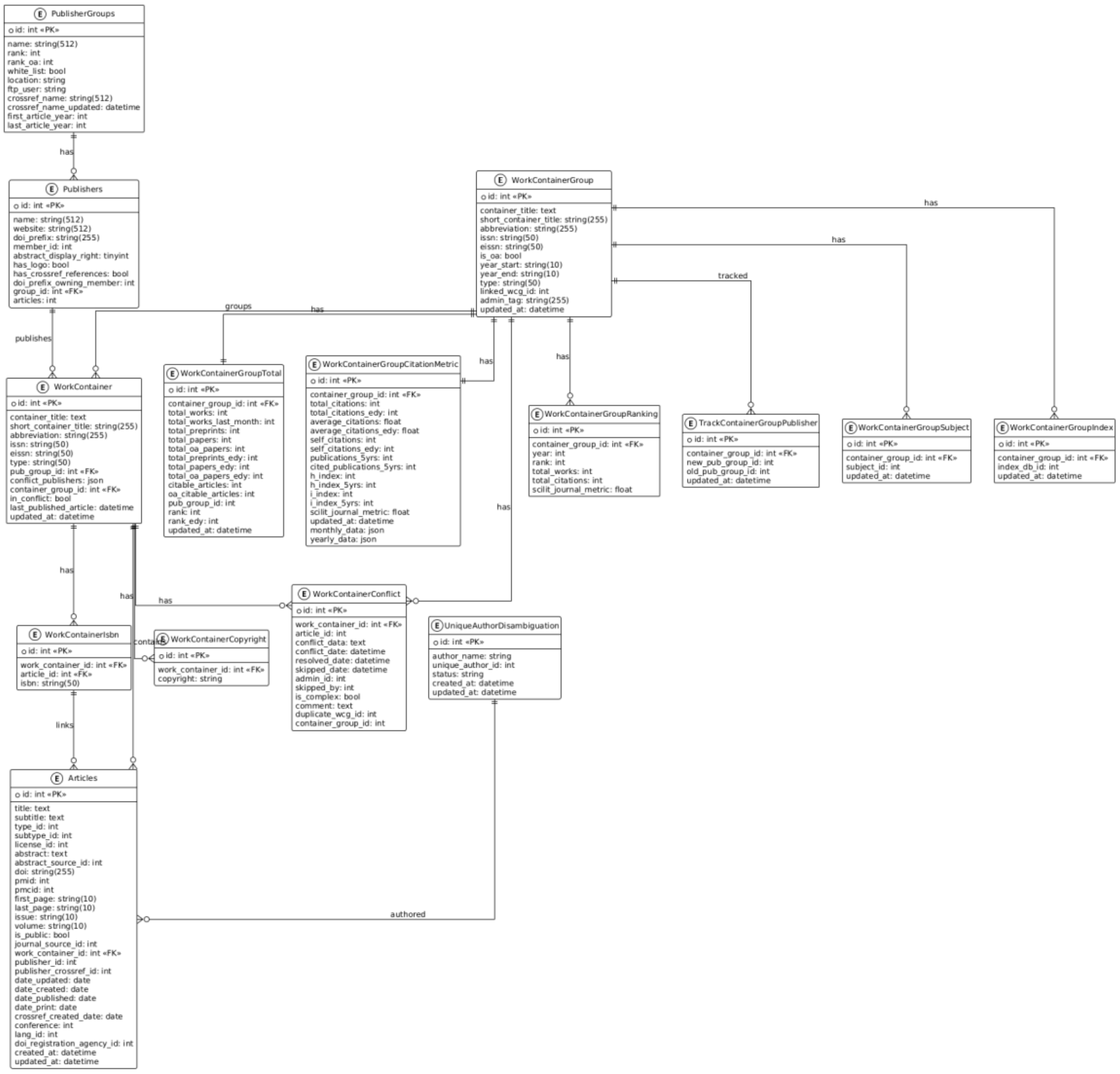

Figure 5. ER diagram of the Source Entity

### *2.7.3 Publishers Entity*

The publisher's system in Scilit is a central place where all information about publishers and publisher groups is collected and managed. Each publisher record stores basic details such as the publisher's name, website, DOI prefix, and several flags that describe how the publisher content should be handled in the platform. Publisher groups act as parent organizations that can include multiple publishers and define who has permissions to manage them.

The system also connects publishers with articles and work containers, making it easier to track where each publication comes from and how it should be indexed. It includes tools for resolving conflicts when different sources provide different data about the same publisher, as well as mechanisms for whitelisting, tracking changes, and updating relationships between publisher groups.

User permissions are an important part of the setup, each user in the system is always associated with one publisher. If a publisher does not yet have a registered user, the user selects the publisher they represent during registration. This association must then be manually reviewed and approved by a Scilit administrator. Once approved, the user becomes linked to that publisher.

There are two user roles: owner and user. The owner role has full management permissions for the publisher, including inviting and removing users, managing access, importing article XML files, and adding or managing other publishers that belong to the same publisher group. The owner can also update the publisher name as it is displayed on Scilit. Regular users have limited permissions and can only perform actions granted by the owner within the scope of the assigned publisher.

Users can own publisher groups, upload files, and request changes depending on their access level. The service layer coordinates all these processes, from validating data and updating records to triggering events when something changes. Overall, the publisher's system provides a clean structure for keeping publisher information consistent, reliable, and well-integrated with the rest of the Scilit indexing workflow.

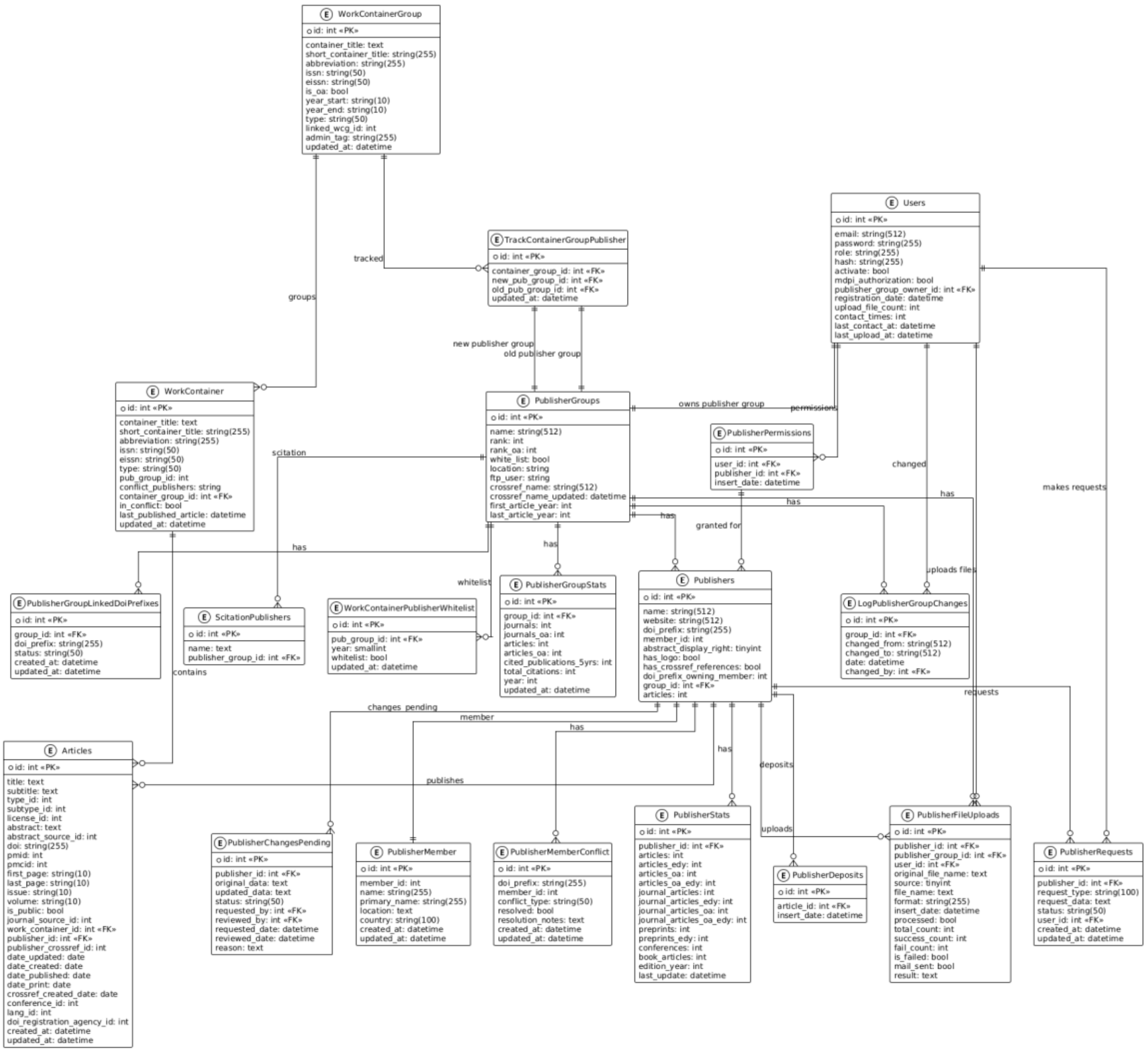

Figure 6. ER diagram of the Publisher Entity

### *2.7.4 Institute Entity*

The Institution System is used to manage and analyze data related to academic and research institutions. It is structured around entities that capture every dimension of institutional management.

The central entity is Institutions which stores key details such as name, country, Grid ID (a global identifier for research institutions), ROR ID (Research Organization Registry ID), established (year founded), and logo URL. It serves as the hub for all institutional data. At the same time, it also uses three Geographic Hierarchy Entities: Countries (stores country names and codes), States (captures state/province information, linked to Countries), and Cities (stores city names, linked to States). In addition, Affiliations is another primary entity which defines specific relationships between authors and institutions. The linked entities are ArticleAuthorsAffiliations (connecting articles, authors, and affiliations) and UniqueAuthorAffiliations (connecting unique authors to affiliations). Regarding the statistics and indicators of Institutions, there are entities InstitutionStats and InstitutionMetrics involved. InstitutionStats provides year-specific statistics, such as highly_cited_scholars, highly_cited_papers, and international_collaboration_rate. InstitutionMetrics captures academic metrics like total_citations_5yrs (number of citations in the past 5 years), and oa_articles (number of open-access articles).

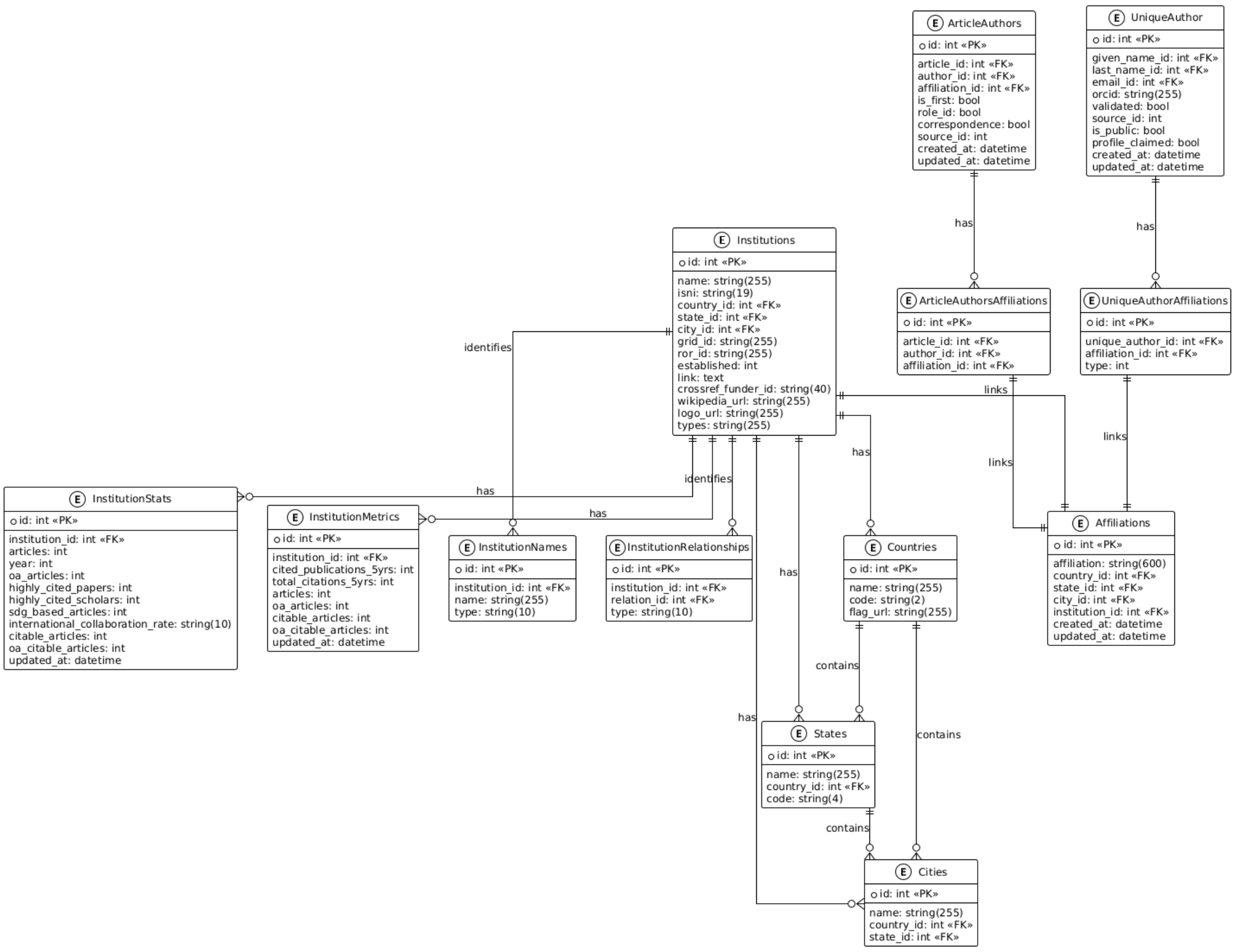

Figure 7. ER diagram of the Institutes Entity

*2.8 Metrics Produced by Scilit*

Scilit provides two-dimensional assessment metrics for quantitative and qualitative aspects as summarized in Table 2. Overall, we employ advanced automated techniques to create four entities: sources, publishers, scholars, and organizations. In each individual entity, users can check the basic information, publishing output in the recent decades, latest publications, and citation metrics in each profile, which helps users to have a clear picture in terms of publishing performance on this entity. This study takes I3 for an absolute perspective and I3/N for a relative perspective as examples for illustration and demonstration.

Table 2. Metrics produced by Scilit.

| Dimension | Metric | Description | Notes / Examples |
|---|---|---|---|
| Output | Total publications | Count of all indexed records over recent decades | Time-trended totals |
| Output | Total open access publications | Count of OA records over recent decades | Time-trended OA coverage |
| Output | Top 10 countries by publications | Publication counts by country across recent decades | Ranked list with counts |
| Output | Top 10 subject areas by publications | Publication counts by subject area across recent decades | Ranked list with counts |
| Impact | Total citations | Aggregate citations received | Basic citation indicator |
| Impact | Average citations | Mean citations per publication | Sensitive to skew; interpret with caution |
| Impact | Self-citations | Self-citation counts across entities | Can be reported by entity type |
| Output | International collaboration rate | Share/level of cross-country coauthorship | Quantifies cross-border collaboration patterns |
| Output | Domestic inter-organizational collaboration | Collaboration across different organizations within a country | Organization-level network intensity |
| Output | Domestic intra-organizational collaboration | Collaboration within the same organization | Internal teaming intensity |
| Impact | Citation Depth | Absolute and relative share of citing publications that also cite the focal paper's references | Reflects deep within-area influence |
| Impact | Citation Breadth | Conceptual breadth across research areas | Captures cross-area reach |
| Impact | Citation Dependence | Absolute: count of citing papers also citing the focal paper's references; Relative: proportion thereof | Indicates reliance on prior work; computed for Top Cited Papers (≥100 citations) |
| Impact | Citation Independence | Absolute: count of citing papers NOT citing the focal paper's references; Relative: proportion thereof | Indicates independent contribution; computed for Top Cited Papers (≥100 citations) |
| Impact | h-index | $h$ publications with $\geq h$ citations | Career-level impact |
| Impact | i10-index | Publications with ≥ 10 citations | Simpler threshold-based impact |
| Impact | H5-index | h-index computed over the last 5 years | Recency-focused journal impact |
| Impact | Monthly Citation Metric | Monthly citation accumulation indicator | Tracks short-term momentum |
| Impact | I3 | Sum of weighted publication counts across citation-percentile classes | Distribution-sensitive and field-comparable; combines volume and impact |
| Impact | I3/N | I3 normalized by publication count | Size-normalized and field-comparable impact measure |

## 3. Data Acquisition and Processing

We calculated the JIF, CiteScore, I3, and I3/n on the journal level for the years 2023 and 2024 as follows:

### *3.1 JIF*

The definition and calculation of JIF (Clarivate, 2025a) are as shown in Equation (1):

$$JIF(Y) = \frac{Citations\ received\ in\ year\ Y\ to\ items\ published\ in\ years\ Y-1\ and\ Y-2}{Number\ of\ citable\ items\ published\ in\ years\ Y-1\ and\ Y-2} \quad (1)$$

where Y represents the JIF year.

We downloaded and obtained the 2023 JIFs for 21,591 journals and the 2024 JIFs for 22,224 journals from the official Journal Citation Reports (JCR) website (https://jcr.clarivate.com). The downloaded data also include information such as journal ISSN/eISSN, publishers, and subject categories, which were used for further data matching and analysis.

### *3.2 CiteScore*

The definition and calculation of CiteScore are shown in Equation (2):

$$CiteScore(Y) = \frac{Citations\ received\ in\ year\ Y\ to\ documents\ published\ in\ years\ Y-1, Y-2, Y-3, and\ Y-4}{Number\ of\ documents\ published\ in\ years\ Y-1, Y-2, Y-3, and\ Y-4} \quad (2)$$

where Y represents the CiteScore year.

We obtained the CiteScore for 2023 for 29,555 journals and 2024 for 30,509 journals through the Scopus API (https://dev.elsevier.com/documentation/ScopusSearchAPI.wadl). The downloaded data also include information such as journal ISSN/eISSN, publishers, and subject categories, which were used for further data matching and analysis.

### *3.3 Scilit I3 and I3/N*

We defined the I3 in Scilit as shown in Equation (3):

$$I3 = \sum_i (x_i \times n(x_i)) \quad (3)$$

where $x_i$ is the weight assigned to percentile rank class $i$ based on papers in top 1%, top 10%, top 50%, and in Bottom 50%; $n(x_i)$ is the number of publications in the percentile class $i$.

The core concept of I3 indicators is to give publications in different citation intervals different weights to appreciate the highly skewed citation distribution (Leydesdorff, Bornmann, & Adams, 2019). The weighting scheme we used for Scilit I3 computation is a bit different from the scheme proposed above. Following the advice of our Advisory Board Members (ABM) of Scilit, we assigned a score of 0 to the bottom 50% papers to adjust the indicator more sensitive to the citation impact side over the output side:

$$Scilit\ I3 = \sum_i (x_i \times n(x_i)) = \sum_{i=1}^{4} w_i \quad (4)$$

Where $w_i$ denotes $x_i \times n(x_i)$. We define four groups of weights $x_1 = 100$, $x_2 = 10$, $x_3 = 2$, $x_4 = 0$, which correspond to citations from the top 1%, top 10%, top 50%, and bottom 50%, respectively. $n(x_i)$ represents the number of papers in the group associated with each weight.

Meanwhile, in determining the time windows for the Scilit I3 definition, we adopted ABM's recommendation by choosing a two-year publication window and a four-year citation window. This ensures that cited publications have a sufficient period to accumulate citations while maintaining the timeliness and informational value of the focal units evaluated by the indicator. In addition, the selection of publication and citation windows in the Scilit I3 indicator also draws on the practices of the (five-year) JIF and CiteScore.

After determining the weight and time windows, for the computation of references in Scilit I3, we took the citing side normalization approach for field normalization (Zhou & Leydesdorff, 2011). For each citation, we calculated the fractional citation count as shown in Equation (5):

$$\text{Fractional citation count} = \sum_i 1/\mathrm{m}_i \quad (5)$$

where $\mathrm{m}_i$ is the number of references in citing article *i*.

We used fractional counting for the following reasons: i) This normalizes for field-dependent citation "potentials" (i.e., different reference-list lengths and citation densities) could reduce cross-field bias without relying on imperfect journal-based field classifications. ii) By normalizing at the level of individual citing papers, fractional counting enables fairer comparisons across disciplines and specialties, and separates size effects (total citations) from relative impact (citations per paper), which is important when aggregating I3 across units. iii) The method provides distributional data (fractional citation weights per item), allowing statistical testing of differences among units, and it has been shown to substantially reduce between-field variance in impact measures and alter rankings in sensible ways (e.g., upgrading of ranks of subjects like Humanities or Engineering where citation densities differ much between fields) (Zhou & Leydesdorff, 2011). Thus, in our opinion, for Scilit I3, which sums weighted contributions across percentile classes or papers, using fractional counts as the citation input could improve cross-disciplinary comparability, additivity, and interpretability of the integrated impact.

After obtaining each paper's total fractional citation count, we rank papers within the same publication year and document type and assign percentile ranks to compute Scilit I3. The Scilit I3 calculation includes the following document types in 2021-2022: research articles (6,569,561), review articles (559,475), conference papers (59,274), case reports (49,272), and clinical trials (353).

To compare with JIF and CiteScore, we defined the Scilit I3/N for the journal level, dividing Scilit I3 by the number of journal publications (N), which expresses the contribution made by an average paper given the journal's characteristics:

$$Scilit\ I3/N = \frac{1}{n_k}\sum_{i=1}^{n_k k} w_i \quad (6)$$

Where $n_k$ is the number of journal publications within the Scilit I3 time window and document type of journal $k$. We calculated Scilit I3 and I3/N for 55,842 journals in 2023 and 61,739 journals in 2024, respectively. The Scilit data also include information such as journal ISSN/eISSN, publishers, and subject categories, which were used for further data matching and analysis.

### *3.4 Data Processing*

A total of 17,816 journals, based on their title, ISSN, or eISSN, were matched for comparison analysis using JIF, CiteScore, Scilit I3, and Scilit I3/N. In order to achieve the comparison of different indicators with the same scale in various dimensions, we used the maximum-minimum normalization method to simultaneously normalize for JIF, CiteSore, Scilit I3, and Scilit I3/N and the number of journal publications within the time window, which are used for the subsequent discussion and analysis. analysis:

$$x' = \frac{x - \min(x)}{\max(x) - \min(x)} \quad (7)$$

where $x$ is one of the journal indicators. We used R (version 4.5.1) (R CoreTeam, 2025) for data matching as well as data visualization.

### *3.5 Retrieval of Subject Classifications and Journal-level Matching*

To ensure that journal attributions from the three databases were comparable within a common subject dimension, we first retrieved the native subject classifications of journals from each source. For WoS, we obtained the Web of Science Categories at the journal level from the JCR database, which provides one or multiple subject categories for each indexed journal.For Scopus, we retrieved journal subject classifications from the Scopus Sources list (Scopus source title list), which assigns one or more All Science Journal Classification (ASJC) codes to each journal. These ASJC codes were then used as the Scopus subject-classification scheme. For Scilit, we used the official subject categories provided at the journal (source) level in the corresponding database documentation or source list as mentioned in Section 2.6. All three sets of subject classifications were matched to our journal universe using ISSN (and e-ISSN when available). When a journal had multiple ISSNs, all valid ISSNs were considered in the matching procedure, and duplicates were removed. Journals assigned to multiple subject categories in a given database were retained with all their subject assignments; our cross-database mapping therefore operates at the level of journal–subject pairs rather than single, mutually exclusive categories. Based on these journal–subject assignments, we then constructed a crosswalk between the three classification systems by identifying journals that appear in more than one database and comparing their assigned categories. The comparative analysis presented in this study is therefore grounded in empirically observed subject assignments of the same journals across WoS, Scopus, and Scilit.

## 4. Results

### *4.1 Statistical Description*

Table 3 reports descriptive statistics for 2023 and 2024, organized by indicator type. Across all major databases, the number of journals covered increased in 2024 compared to 2023. Notably, the number of journals with Scilit I3 (from Scilit) is roughly twice that of journals with CiteScore (from Scopus), highlighting the coverage advantage of non-exclusive databases.

For JIF, 2024 shows a modest rise in central tendency accompanied by a marked reduction in extremes, indicating a slightly more concentrated distribution. By contrast, CiteScore remains broadly stable across the two years in terms of mean and median, but exhibits greater dispersion, suggesting that top-end CiteScores rose further.

Compared with 2023, Scilit I3/N in 2024 declines in both mean and median while its dispersion increases, pointing to widening variation within the journal population. For Scilit I3, the overall mean is essentially unchanged, but the median decreases and the upper tail extend, indicating stronger extremes despite a stable average.

Table 3. Descriptive statistics for indicators.

| Indicator | Year | N | Min | Mean | Median | Max | 95% CI | SD | SE |
|---|---|---|---|---|---|---|---|---|---|
| JIF | 2023 | 21,951 | 0 | 2.27 | 1.4 | 521.60 | [2.20, 2.34] | 5.19 | 0.04 |
| JIF | 2024 | 22,224 | 0 | 2.37 | 1.5 | 232.40 | [2.31, 2.42] | 4.26 | 0.03 |
| CiteScore | 2023 | 29,555 | 0 | 3.64 | 2 | 873.20 | [3.55, 3.72] | 7.57 | 0.04 |
| CiteScore | 2024 | 30,509 | 0 | 3.64 | 2 | 1,154.20 | [3.54, 3.73] | 8.68 | 0.05 |
| Scilit I3/N | 2023 | 55,842 | 0 | 1.39 | 0.76 | 100.00 | [1.36, 1.41] | 2.91 | 0.01 |
| Scilit I3/N | 2024 | 61,739 | 0 | 1.34 | 0.67 | 100.00 | [1.32, 1.37] | 3.17 | 0.01 |
| Scilit I3 | 2023 | 55,842 | 0 | 246.94 | 12 | 161,566.00 | [230.55, 263.32] | 1,975.16 | 8.36 |
| Scilit I3 | 2024 | 61,739 | 0 | 244.89 | 10 | 183,360.00 | [228.67, 261.11] | 2,056.77 | 8.28 |

*4.2 Comparison between JIF, CiteScore, and Scilit I3(/N)*

*4.2.1 Indicator Distribution*

Figure 8 displays scatterplots for the three pairs of indicators and annotates the corresponding coefficients of determination ($R^2$) for the fitted curves, Spearman rank correlations, and Lin's concordance correlation coefficients (CCC). Unlike the Spearman rank correlation, Lin's concordance correlation coefficient is not a measure of correlation but agreement. The two-year distributions for all three pairs appear broadly similar, and overall performance is consistently strong. In terms of correlation and concordance, the ordering is approximately CiteScore ≳ I3/N > Journal Impact Factor, which may reflect the wider citation window used by CiteScore, leading to reduced temporal volatility.

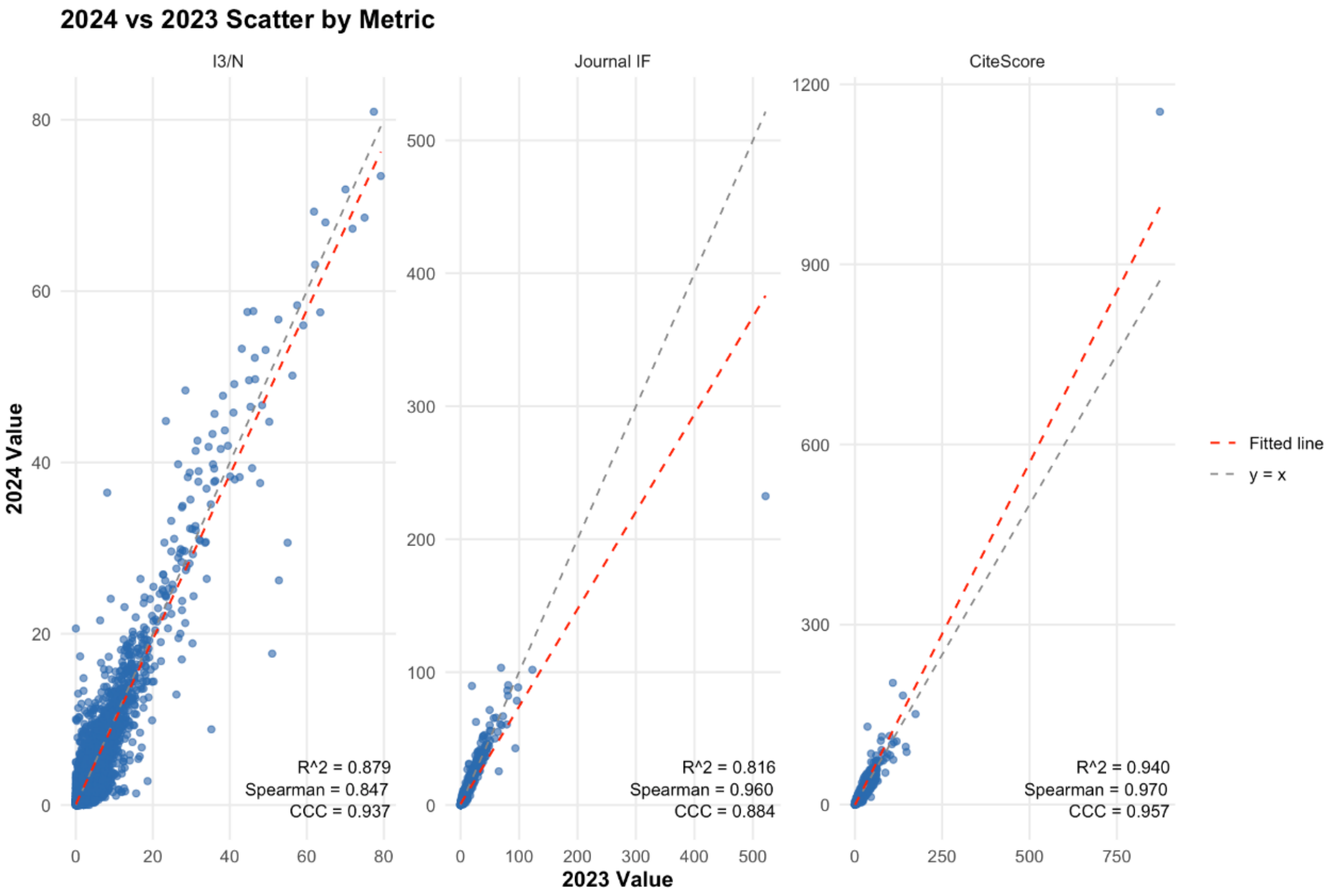

Figure 8. 2023–2024 Scatter Comparison of I3/N, JIF, and CiteScore.

Table 4 assesses both the correlation and agreement among the three indicators. Despite a marked drop in overall agreement in 2024 (CCC falling from 0.81 to 0.56), JIF retains a very high rank correlation with CiteScore. By contrast, Scilit I3/N and JIF/CS exhibit only moderate correlations, indicating that I3/N captures structural impact distinct from average citation-based indicators. Notably, the scales of I3/N and JIF converge in 2024 (CCC rises to 0.72), likely because the two series share more similar trend; however, the gap with CiteScore widens (CCC declines to 0.30), which aligns with the fitted-curve patterns in Figure 8.

Table 4. Correlation and Concordance among I3/N, JIF, and CiteScore (CS)

| Year | Pair | Spearman | CCC |
|---|---|---|---|
| 2023 | I3/N & CS | 0.65 | 0.36 |
| 2023 | I3/N & JIF | 0.64 | 0.54 |
| 2023 | JIF & CS | 0.94 | 0.81 |
| 2024 | I3/N & CS | 0.62 | 0.30 |
| 2024 | I3/N & JIF | 0.62 | 0.72 |
| 2024 | JIF & CS | 0.94 | 0.56 |

Figure 9 presents the empirical cumulative distribution functions (ECDFs) for the three indicators and contrasts their two-year dynamics. In Figure 9A, the ECDF for I3/N is smoother, strictly monotonically increasing, and exhibits fewer inflection points, indicating a more continuous distribution without pronounced discontinuities. The distribution of I3/N also appears more compact, with a shorter span along the horizontal axis, indicating compressed cross-journal heterogeneity. By contrast, the JIF ECDF is more segmented, with locally steeper increments that imply clustering of journals within specific value ranges; its upper tail is longer yet constitutes a relatively small share of the total number of journals. The CiteScore ECDF is smoother than that of JIF, plausibly owing to its longer aggregation window and larger denominator. Figure 9B further indicates that the two-year distribution for I3/N is most salient between the low and the medium-high ranges, potentially reflecting heightened sensitivity of this metric to shifts in citation structure or to its normalization scheme. Around the median, the negative differential for JIF is consistent with a modest rightward displacement in 2024, suggesting a slight, albeit not substantial, increase in overall journal impact. By comparison, CiteScore remains largely stable, implying limited year-over-year variability in its computational design or effective sample size.

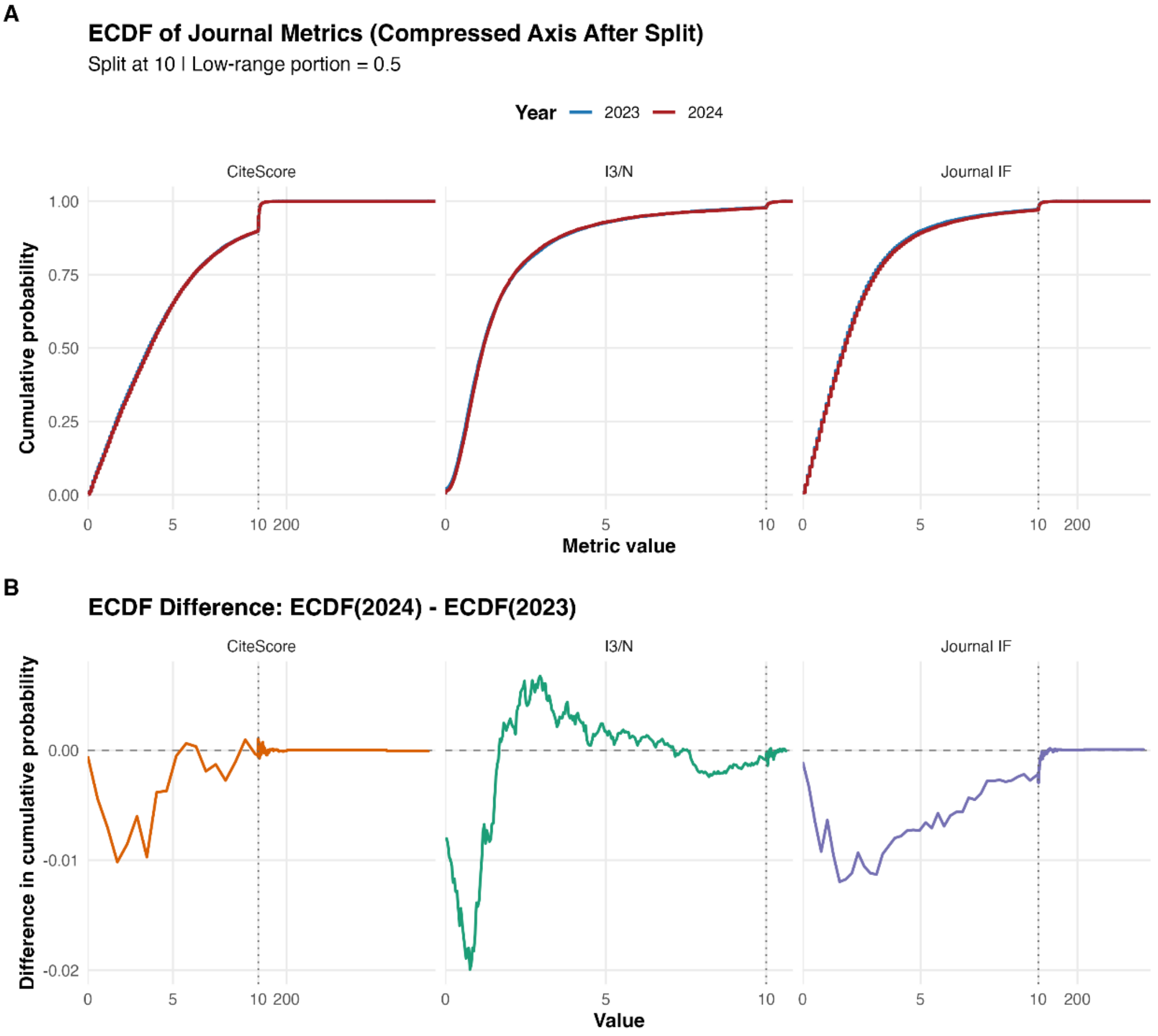

Figure 9. Distribution of journal metrics for empirical cumulative distribution function (ECDF) comparisons with annual differentials: A) the ECDF of CiteScore, Scilit I3/N, and JIF for 2023 vs. 2024, with the horizontal axis compressed after a value of 10 to emphasize the details of the low-middle range, and B) the ECDF difference of 2024 relative to 2023, i.e., ECDF(2024) - ECDF(2023), which is used to determine the left or right shift of the distribution.

#### *4.2.2. Quadrant-Based Assessment*

Conceptually, Scilit I3 is a "scale-sensitive" total-impact metric that reflects publication volume and output size, whereas JIF, CiteScore, and I3/N function as "density" metrics that attenuate size effects. These density metrics, however, display distinct distributional dynamics across years. Taken together, Scilit I3 and I3/N from the Scilit database provide a more comprehensive assessment of journals and help reveal divergences between scale effects and unit-quality performance.

Figure 10 evaluates journals' scale effect and unit quality using Scilit I3 and Scilit I3/N, and visualizes matched journals for 2023 and 2024 (Figure 10A–B). In Figure 10A, journals in Quadrant 1 (upper right, e.g., *Nature Communications*, *Nature*, *Science*) exhibit simultaneously high I3 (overall impact magnitude) and high I3/N (impact per publication), indicating that these outlets are both "large" and "strong." Quadrant 2 journals (e.g., *PLOS ONE*, *Scientific Reports*) show reasonable performance in I3 (i.e., a noticeable scale effect) but comparatively limited Scilit I3/N, suggesting that unit quality is less pronounced than in top-tier venues. Such journals are often multidisciplinary or publish substantial volumes within specific research field (e.g., *Sustainability* and *Science of the Total Environment* in environmental science). Quadrant 4 contains clusters of prominent medical journals, reflecting high unit quality accompanied foremost by substantial publication volume. Figure 10B displays the 2024 distribution of Scilit I3 and Scilit I3/N, which broadly mirrors the 2023 pattern; however, the relative positions of high-impact journals in Quadrant 1 appear slightly more separated, implying a potential further concentration among headline journals.

To further elucidate the properties of Scilit I3 and Scilit I3/N, we standardized Scilit I3, Scilit I3/N, CiteScore, and publications within a time window and compared the Scilit-based metrics with the standardized CiteScore and publications along two dimensions: scale effect and unit quality, to assess their distinctive characteristics. Results from the two-year comparison are shown in Figure 10C–D. As illustrated in Figure 10C, journals in the first quadrant (Unit Quality > 0 and Scale Effect > 0) exhibit advantages in both dimensions relative to the reference (CiteScore and publication volume). Representative journals include *Nature Communications*, *Nature*, *Science*, and *IEEE Access*, indicating that I3/N captures a unit-quality advantage over CiteScore, while I3 suggests that an equivalent number of publications can yield greater overall impact.

By contrast, the second quadrant (Unit Quality < 0 but Scale Effect > 0) reflects strong scale effects but no unit-quality superiority relative to CiteScore. This represents journals typically with large publication volumes and substantial aggregated citations but comparatively limited per-article performance. The most salient example is *CA: A Cancer Journal for Clinicians*. Notably, although this journal ranks first in the 2023 CiteScore list and its Scilit I3 exceeds its publication volume, indicating that scale effects render Scilit I3 a more objective indicator of impact relative to publication volume. On the other hand, its Scilit I3/N substantially diminishes the corresponding CiteScore-based impression of unit quality. In other words, I3/N attenuates the long-tail effect and provides a more balanced assessment of per-article quality.

Finally, journals in the fourth quadrant (Unit Quality > 0 but Scale Effect < 0), such as *PLOS ONE*, *Energies*, and *Sustainability*, display an advantage in unit quality but an insufficient scale effect, i.e., overall impact that is modest relative to their publication counts.

Compared to 2023, the representative journals in the field of engineering and environment, *IEEE Access* and *Journal of Cleaner Production*, show a more pronounced improvement in the scale effect. Surprisingly. the *International Journal of Systematic and Evolutionary Microbiology* journal excels in the unit quality dimension. This might be an artifact of the fractional citation counting that the fractional count for most of the journal's citations is calculated as 1 due to incomplete metadata for journal citations. As a result, , the journal impact is over-represented under the Scilit I3 and I3/N.

Quadrant assessments based on the Scilit database provide complementary perspectives on quality per unit (I3/N) and scale effects (I3). While some journals have an average quality advantage through high I3/N, others demonstrate aggregation of impact at scale through high I3, and the top journals in Quadrant 1 have both. In summary, the Scilit I3-I3/N four-quadrant framework provides a more granular and diagnostic-friendly view of journal impact, complementing the explanatory power of the combination of publications and CiteScore.

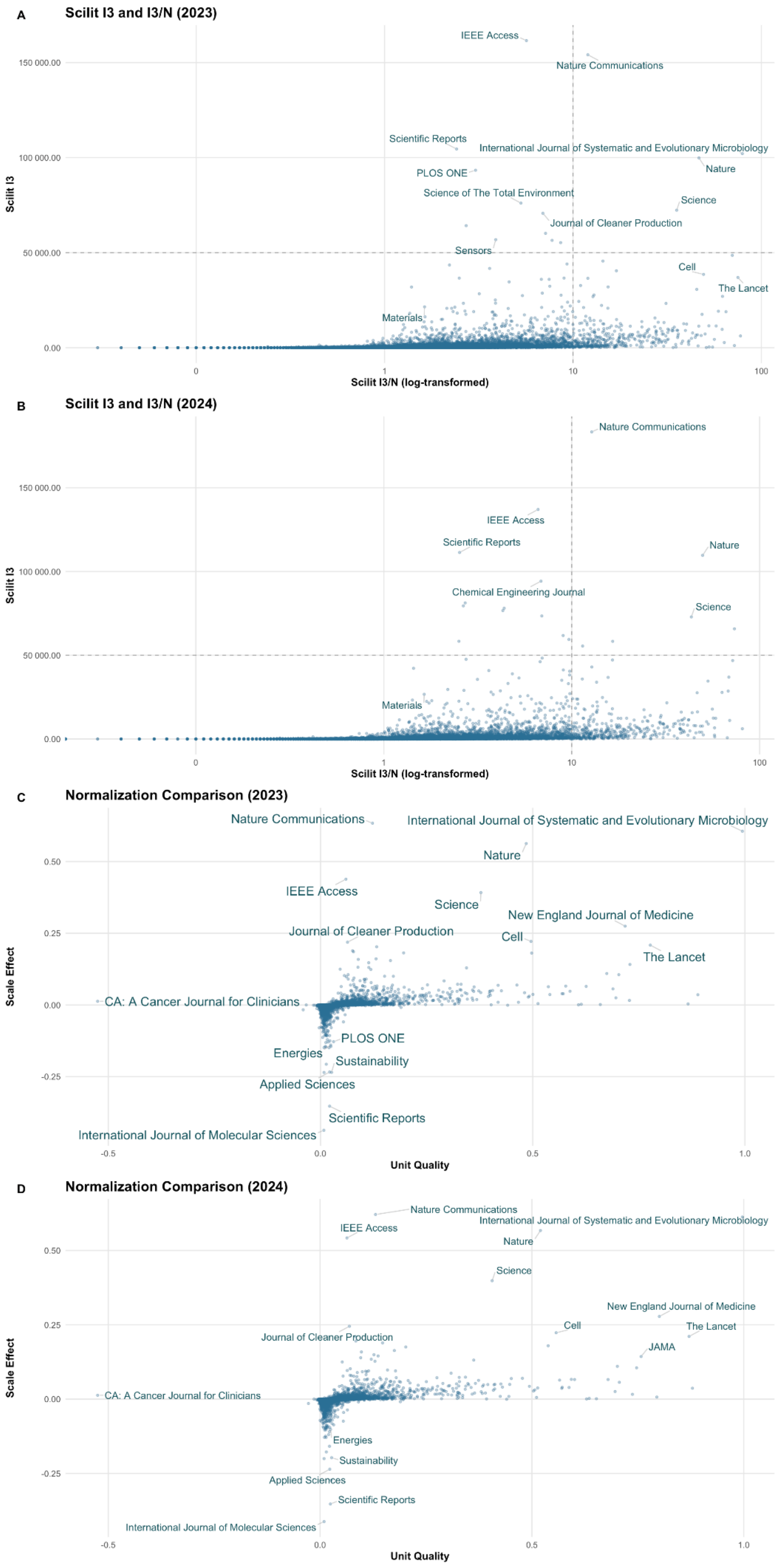
A
Scilit I3 and I3/N (2023)
Scilit I3
Scilit I3/N (log-transformed)
150 000.00
100 000.00
50 000.00
0.00
0
1
10
100
IEEE Access
Nature Communications
Scientific Reports
International Journal of Systematic and Evolutionary Microbiology
Nature
PLOS ONE
Science of The Total Environment
Science
Journal of Cleaner Production
Sensors
Cell
The Lancet
Materials
B
Scilit I3 and I3/N (2024)
Scilit I3
Scilit I3/N (log-transformed)
150 000.00
100 000.00
50 000.00
0.00
0
1
10
100
Nature Communications
IEEE Access
Scientific Reports
Nature
Chemical Engineering Journal
Science
Materials
C
Normalization Comparison (2023)
Scale Effect
Unit Quality
0.50
0.25
0.00
-0.25
-0.5
0.0
0.5
1.0
Nature Communications
International Journal of Systematic and Evolutionary Microbiology
Nature
IEEE Access
Science
New England Journal of Medicine
Journal of Cleaner Production
Cell
The Lancet
CA: A Cancer Journal for Clinicians
PLOS ONE
Energies
Sustainability
Applied Sciences
Scientific Reports
International Journal of Molecular Sciences
D
Normalization Comparison (2024)
Scale Effect
Unit Quality
0.50
0.25
0.00
-0.25
-0.5
0.0
0.5
1.0
Nature Communications
International Journal of Systematic and Evolutionary Microbiology
IEEE Access
Nature
Science
New England Journal of Medicine
Cell
The Lancet
Journal of Cleaner Production
JAMA
CA: A Cancer Journal for Clinicians
Energies
Sustainability
Applied Sciences
Scientific Reports
International Journal of Molecular Sciences

Figure 10. Quadrant-based assessment: A) Visualization of matched journals' Scilit I3 and I3/N based on 2023; B) Visualization of matched journals' Scilit I3 and I3/N based on 2024; C) Normalization Comparison of Scilit I3 and I3/N with CiteScore and Journal Publication (2023) and D) Normalization Comparison of Scilit I3 and I3/N with CiteScore and Journal Publication (2024).

#### *4.2.3. Heterogeneity Across Databases Within Disciplines*

To make journal attributions across the three databases comparable within a common subject dimension, we mapped the three subject-classification systems using the subject assignments of the same journals in each database and conducted a cross-database comparative analysis accordingly. Figure 11 shows the cross-database co-occurrence network of subjects (subject IDs) under the threshold of more than 200 overlapping journals. Nodes represent subject IDs. A strong correspondence among the three databases' discipline definitions can be identified when two or more differently colored edges converge on the same node. For example, the triangular connections among 76 (Scilit: Economics), 386 (Scopus: Economics and Econometrics), and 648 (WoS: Economics) indicate substantial overlap in economics-related journals, suggesting a consistent cross-database alignment despite differences in naming and granularity. Additionally, the relatively thicker blue edges point to greater overlap between WoS and Scopus in certain fields, while the distribution of green edges highlights areas where Scilit aligns with the other two databases. Overall, this figure reveals a strong co-occurrence and substantial overlap across several key disciplines, providing a foundation for subsequent standardized mapping, discrepancy analysis, and metric alignment at the disciplinary level.

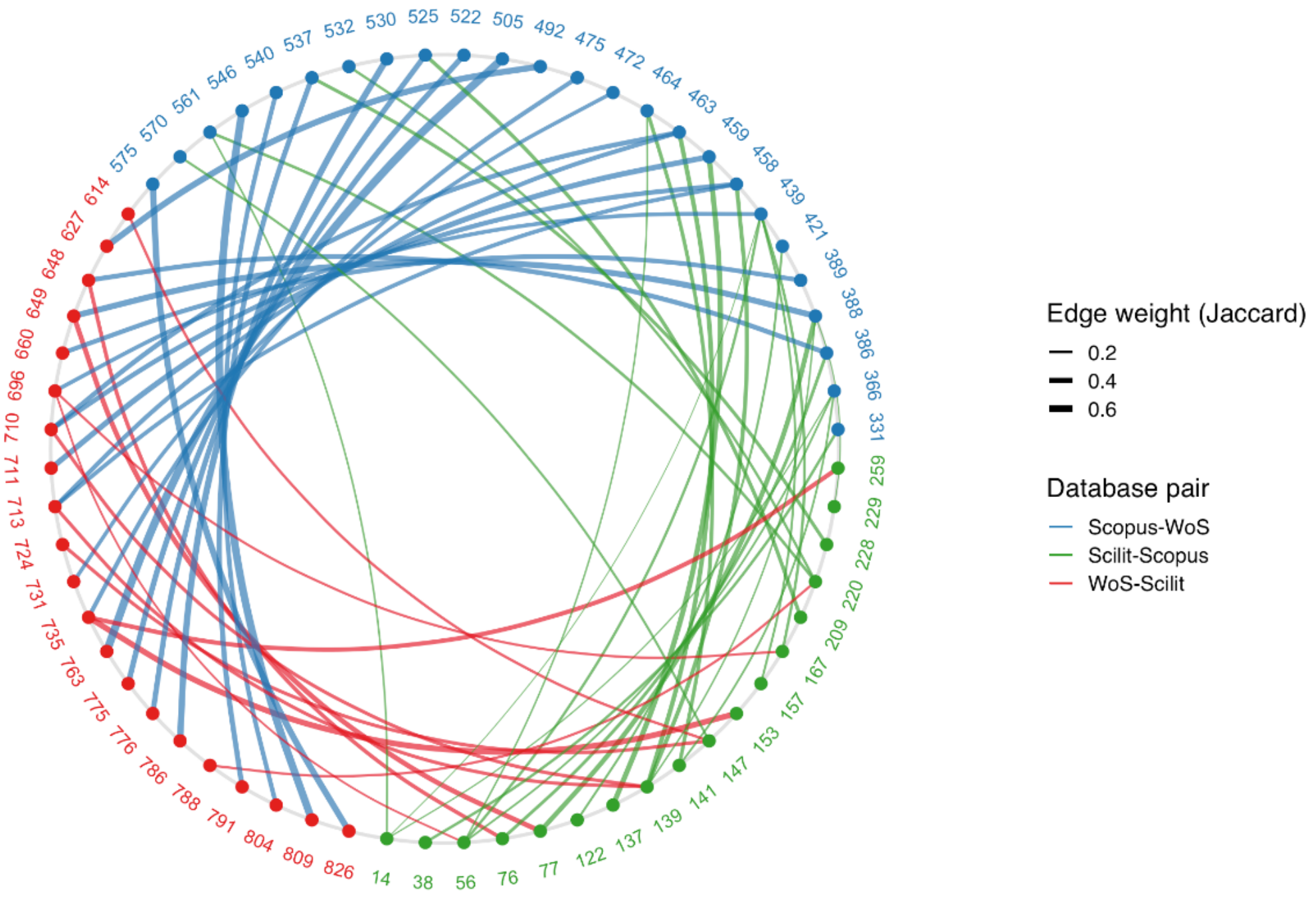

Figure 11. Cross-database subject co-occurrence with journal overlap greater than 200. Each point represents a subject ID: 1–290 = Scilit Subjects; 291–587 = Scopus Subjects; 588–841 = WoS Categories.

After establishing the mapping of discipline definitions across databases, we examined the performance of each set of indicators for the paired journals in 2024 compared to 2023. Table 5 lists some representative subject areas and indicator characteristics, reflecting some commonalities and differences between different databases and different indicators. In the humanities and social sciences (e.g., linguistics, law) and some engineering and materials fields (e.g., energy, electrical and electronics, and metallurgy), all three indicators show significant improvement in 2024. This suggests that the citation ecology, platform coverage, and traditional influence of a journal cluster will improve when the discipline is in a period of steady growth. Certain subfields of life medicine (e.g., Infectious Diseases) reflect a significant downward trend in all three groups of indicators when the topic fluctuates and the manuscript source structure is adjusted, reflecting that changes in the manuscript source structure in the post-COVID phase and the adjustment of journal strategies have characterized the impact of the field as weakening. In the frontier direction of AI, Scilit's I3/N and CiteScore tend to reflect heat and citation activity first, while JIF's feedback lags. In the cross-database mapping of pedagogy, WoS's JIF and Scopus's CiteScore show an upward trend, while Scilit's I3/N shows a downward trend implying that the header and mean quality improve, but the diffusion on the platform side decreases; on the other hand, the number of journals in this field in the scilit database is very prominent, with an increase of 488 journals in 2024 compared to 2023, which affects the distribution of the original journal data to a certain extent. In summary, JIF anchors quality and prestige, CiteScore measures broad-spectrum mean, and I3/N captures diffusion rate. Interpreting the three in a mapped disciplinary classification can effectively avoid the misjudgment of a single indicator, thus forming a complete view of disciplinary growth.

Table 5 Representative cross-database subject mapping with overlapping journals and trend signals. ↑*/↓* indicate a significant upward (downward) trend in 2024 relative to 2023; — indicates no significant change.

| Data subject mapping | Overlapping Journals | Scilit I3/N | JIF | CiteScore |
|---|---|---|---|---|
| Scopus: Education<br>WoS: Education & Educational Research<br>Scilit: Education & Pedagogy | 405 | ↓* | ↑*** | ↑* |
| Scopus: Economics and Econometrics<br>WoS: Economics<br>Scilit: Economics | 283 | — | — | — |
| Scopus: Linguistics and Language<br>WoS: Language & Linguistics<br>Scilit: Linguistics & Language Studies | 214 | ↑* | ↑*** | ↑*** |
| Scopus: Law<br>WoS: Law<br>Scilit: Political Science & International Relations | 108 | ↑** | ↑*** | ↑*** |
| Scopus: Infectious Diseases<br>WoS: Infectious Diseases<br>Scilit: Virology | 64 | ↓*** | ↓*** | ↓*** |
| Scopus: Mechanics of Materials<br>WoS: Materials Science, Multidisciplinary<br>Scilit: Metallurgical Engineering | 62 | ↑* | ↑*** | ↑** |
| Scopus: Electrical and Electronic Engineering<br>WoS: Engineering, Electrical & Electronic<br>Scilit: Cybersecurity | 46 | ↑* | ↑* | ↑* |
| Scopus: Energy Engineering and Power Technology<br>WoS: Energy & Fuels<br>Scilit: Power Systems & Electric Vehicles | 33 | ↑* | ↑* | ↑** |
| Scopus: Artificial Intelligence<br>WoS: Computer Science, Artificial Intelligence<br>Scilit: AI & Machine Learning | 61 | ↑*** | — | ↑** |

*4.2.4. Discrepancies in I3/N and CiteScores*

To examine to what extent the proposed I3/n indicator would induce changes in journal assessments, we investigate the discrepancies in the percentile ranks of the journals' I3/n indicators and CiteScore indicators. For each journal that we have access to both its I3/n indicator and its CiteScore indicator, we calculate the journal's percentile ranks of those indicators and take a difference between the two values. We follow the formula below to do the calculation:

$$rank_difference_i = rank_i3n_i - rank_citescore_i \qquad (8)$$

Where $rank_difference_i$ is the percentile rank difference of journal $i$, $rank_i3n_i$ and $rank_citescore_i$ are the I3/n percentile rank and the CiteScore percentile rank of journal $i$.

We sort the journals in the ascending order of their $rank_difference_i$ and divide the journals into quartiles, thereby the journals in quartile 1 to quartile 4 vary from low average $rank_difference_i$ to high average $rank_difference_i$. Table 6 presents a few summary statistics of the journal quartiles.

Table 6. Summary statistics of the journal groups.

| Group | (a) Average scholarly outputs | (b) Average citation counts | (c) Average I3/n | (d) Average CiteScore | (e) Average percent rank (I3/n) | (f) Average percent rank (CiteScore) |
|---|---|---|---|---|---|---|
| 1 | 445.98 | 2697.4 | 0.7 | 4.43 | 0.31 | 0.61 |
| 2 | 542.19 | 4405.8 | 1.29 | 4.81 | 0.45 | 0.54 |
| 3 | 571.92 | 5402.89 | 3.32 | 6.18 | 0.59 | 0.55 |
| 4 | 260.05 | 581.04 | 2.09 | 1.54 | 0.64 | 0.28 |

In this table, we divide the 24,687 journals into quartiles based on their $rank_difference_i$ which are computed based on equation (8):

$$rank_difference_i = rank_i3n_i - rank_citescore_i$$

Where $rank_difference_i$ is the percentile rank difference of journal $i$, $rank_i3n_i$ and $rank_citescore_i$ are the I3/n percentile rank and the CiteScore percentile rank of journal $i$.

Table 6 shows that the average percent ranks of I3/n of quartile 1 are evidently lower than that of CiteScore, while that order is clearly reversed in quartile 4. This evidence is not surprising because according to the journal grouping method, quartile 1 should be the journals that in general have a significant downgrade in journal rankings based on I3/n, whereas group 4 should contain the journals whose ranks are markedly increased relative to their ranks of CiteScore after I3/n is introduced.
In terms of number of scholarly outputs, and citation counts, journals in quartile 4 are remarkably lower than the journals in the other groups despite the fact that the quartile 4 journals are the beneficiaries of the shift from evaluations based on CiteScore to evaluations based on I3/n. An important takeaway from the evidence that quartile 4 does not have high numbers of scholarly outputs is that numbers of scholarly outputs which are demonstrated to be a key driver of variations in I3 have been to a very large extent controlled for by the I3/n metric. We attribute this evidence to the possibility that those beneficiary journals are lowly graded in the CiteScore metric due to the low number of citations gained by the journals, but those journals have a high proportion of highly cited publications which are emphasized in the I3/n metric. This inference is also verified by column (c) and column (d). Those two columns indicate that the beneficiary journals on average have low CiteScore values, which are a citation-number-based measure adjusting for numbers of scholarly outputs, compared with the journals in the other groups, but the beneficiary journals outperform the journals in group 1 and group 2 in terms of I3/n, which are mainly a citation-rank-based measure after numbers of scholarly outputs are accounted for.

Furthermore, we examine whether the introduction of I3/n has an uneven impact on journals of different subjects. To alleviate the bias that could be driven by the low numbers of scholarly outputs of some subjects, we only consider the subjects whose numbers of scholarly outputs rank in the first quartile of the entire subject sample set. For each CiteScore subject, we compute the number of journals in each abovementioned journal quartile. We then compute the proportion of number of journals in each quartile for each subject. We first sort the subjects by the proportion of number of journals in quartile 1 in the descending order, then we carry out this method for quartile 4. For display, we only present the top ten subjects in the orderings. The results are shown in Figure 12.

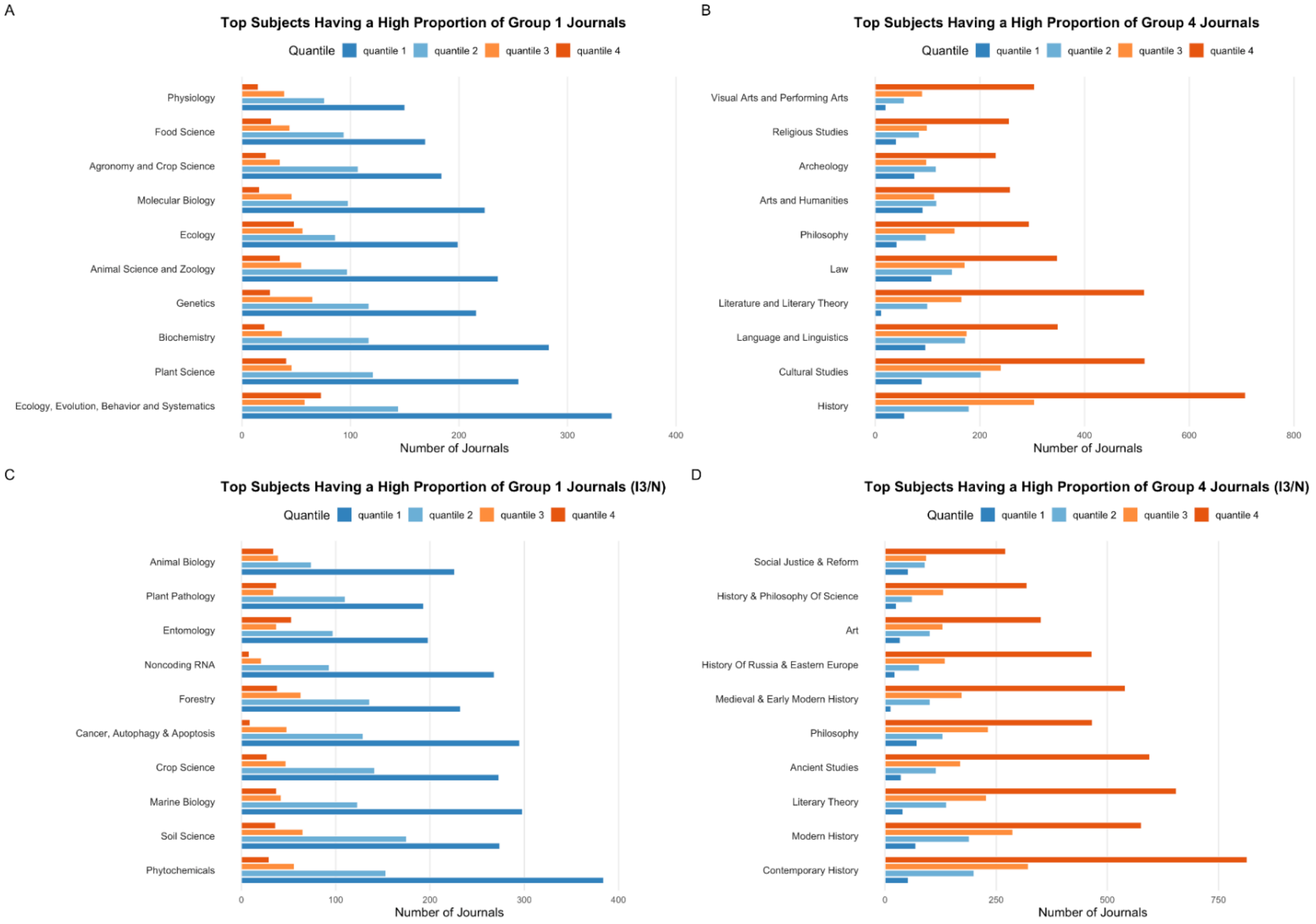

Figure 12. Top Subjects by High Proportions of Journals Within Different quartiles in CiteScore and Scilit.

We count the number of journals within each subject and sort the subjects in the descending order based on the numbers of journals and select the subjects whose numbers of journals rank in the 25 percentile to ensure sufficient data representation. We next sorted the subjects based on their proportions of journals in quartile 1 (sub-figure A) and quartile 4 (sub-figure B). We selected the top ten subjects that have the highest proportions in the investigated quartiles for displaying. The x-axis shows the number of journals covered in each quartile and the y-axis shows the subjects. We also display the number of journals in each quartile in the multiple bars. Sub-figure C and sub-figure D are the analogy of Scilit subjects.

Sub-figure 12A shows that the subjects having a large proportion of journals that would be significantly downgraded by I3/n concentrate on biology, ecology, agriculture, and so on. By contrast, sub-figure 12B implies that the subjects related to the Humanities greatly benefit from the evaluation change from CiteScore to I3/n. The subjects involved in sub-figure 12A and sub-figure 12B have a difference in citation numbers and scholarly outputs. Sub-figure 12C and sub-figure 12D show the analogous analyzes regarding the subject classifications of Scilit. Those subjects that would have strong declines in journal ranking after the introduction of I3/n are concerned with biology, ecology, agriculture, and animal science, while the journals in the subjects that would outperform the other subjects under the evaluative framework of I3/n are related to history, literacy, art, and philosophy. The patterns displayed in these two sub-figures are generally consistent with sub-figure 12A and sub-figure 12B. Therefore, the introduction of I3/n would have uneven impacts on subjects of different citation cultures and publishing frequencies, and the impacts of research related to social sciences could be previously underestimated in the CiteScore metric but more fairly evaluated by the I3/n metric.

### *4.3 Analysis on the Basis of Publishers*

Figure 13 shows the distribution of I3/N for the last two years for journals from major publishers, and uses the relative amount of change (Δ) in the two-year median difference of this indicator and with the Wilcoxon test to mark significance (e.g., “Stable”, “Increase”, or “Decrease”). It is easy to see that the distribution of mainstream general publishers (e.g., Elsevier, Springer Nature, Wiley, SAGE) shows a typical pattern of concentration at the low end and lengthening of the upper tail, with the overall annual median change close to zero, indicating that the portfolio quality has significant inertia and robustness in large samples. In contrast, Frontiers Media SA exhibits the most pronounced downward shift in median (~-11.5%) alongside journal volume expansion, while Wolters Kluwer Health and Oxford University Press display a moderate downward shift, and MDPI and American Chemical Society remain largely stable. This evidence suggests that, on the one hand, journal volume is positively correlated with stability, and the larger the size, the less likely the median is to be shifted by annual perturbations; on the other hand, rapid expansion may bring about a "dilution effect", in which the limited influence of newly added journals offsets the impact contributed by incumbent titles, thereby pulling down the publisher-level distribution.

Combined I3/N distributions from the perspective of publishers, our results indicate that portfolios with broad, established coverage tend to maintain a stable central tendency year over year, even as tail behavior varies. When expansion is rapid and concentrated in newer or lower-impact titles, the median I3/N may decline despite growth in total number of journals, highlighting a trade-off between scale and short-term impact consolidation. Accordingly, academic publishers can use annual I3/N indicators from the Scilit database, together with journal counts, to inform policies on structural improvements and journal release.

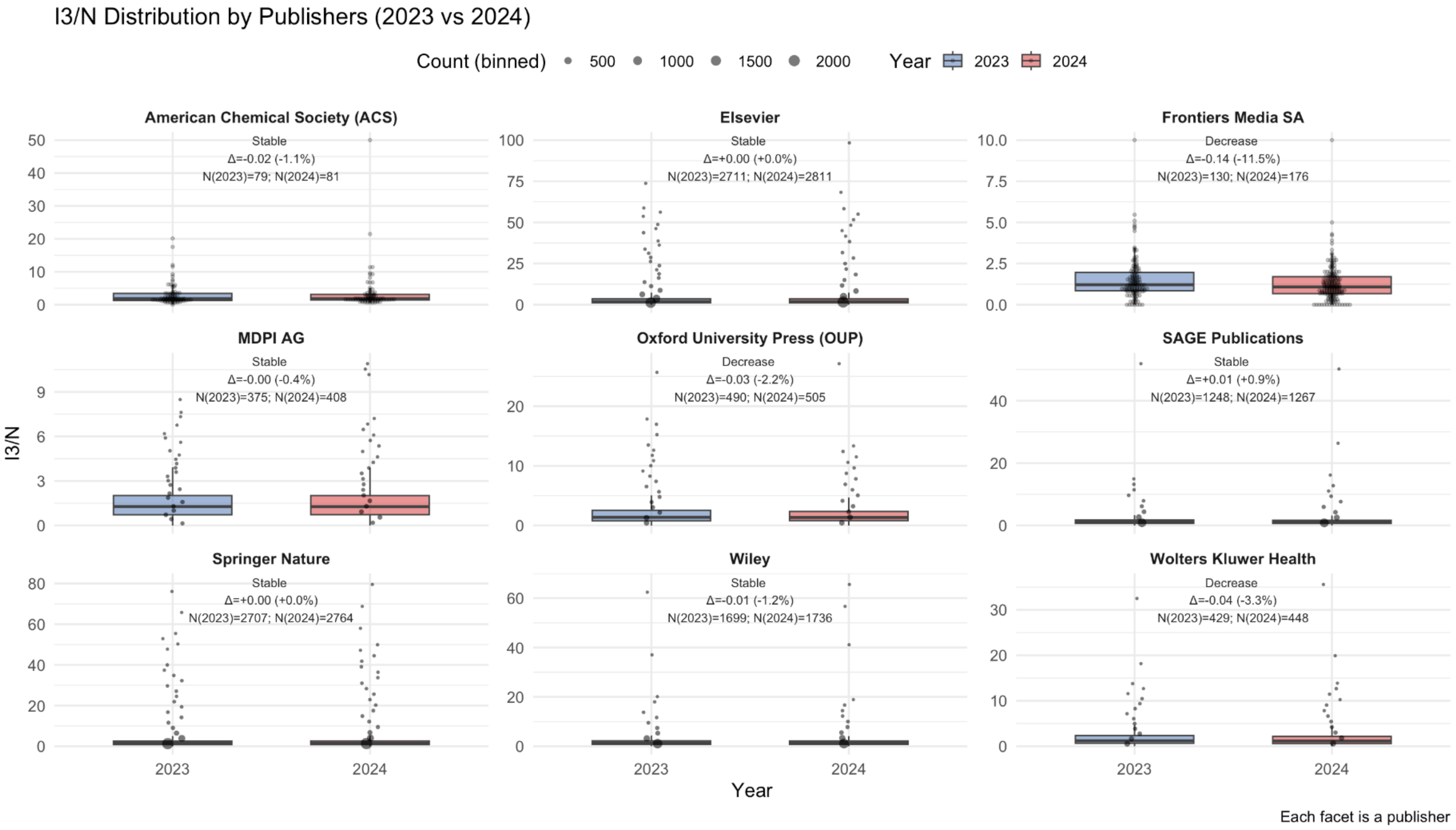

Figure 13. I3/N distribution by top publishers. Δ represents the relative percentage of median difference with the Wilcoxon test result (“Stable”, “Increase”, or “Decrease”), and N represents the total count of journals for each publisher in that year.

## 5. Discussion

This study set out to introduce the Scilit database as a comprehensive, non-exclusionary scholarly platform and to implement the Integrated Impact Indicator (I3) within its framework as a robust alternative for journal impact assessment. Our analysis, benchmarking Scilit, I3, and its normalized counterpart (I3/N) against the established metrics of JIF and CiteScore, yields several critical insights. The central finding here in the current paper is that the combination of Scilit's expansive coverage and the I3's distribution-sensitive methodology offers a more nuanced, equitable, and diagnostically powerful framework for evaluating journal influence, particularly in cross-disciplinary contexts.

The most immediate advantage demonstrated by our results is the dramatic increase in journal coverage. The fact that Scilit computes I3 metrics for over 60K journals, approximately double the number covered by Scopus's CiteScore, is not merely a quantitative achievement but also a qualitative one. Scilit directly addresses a fundamental critique of traditional databases: Their inherent coverage biases toward English-language journals from large commercial publishers, often at the expense of regional, society-owned, or smaller open-access outlets (Larivière, Haustein, & Mongeon, 2015). By leveraging a multi-source ingestion model, Scilit mitigates these biases, thereby democratizing the visibility of a wider spectrum of scholarly work. This expansive coverage is a prerequisite for a more inclusive and globally representative bibliometric analysis.

Methodologically, the I3 framework proves its value by fundamentally rethinking how impact is aggregated. Unlike JIF and CiteScore, which are simple averages and thus susceptible to distortions by a few highly cited articles and the problematic numerator-denominator asymmetry, I3 accounts for the full citation distribution. Our implementation, which assigns weights of 100, 10, 2, and 0 to publications in the top 1%, 10%, 50%, and bottom 50% percentiles respectively, deliberately shifts focus toward high-impact research while effectively ignoring uncited or minimally cited papers. This approach is statistically more robust, as it is less vulnerable to skew and strategic manipulation through citation stacking. The quadrant analysis (Figure 10) powerfully illustrates the diagnostic capability of this dual-metric system. It allows for a clear distinction between journals that are "large and strong" (high I3 and I3/N, e.g., *Nature*), those that are "large but diffuse" (high I3, low I3/N, e.g., PLOS ONE), and those that are "focused and high-quality" (low I3, high I3/N). This disaggregation of scale effect from unit quality provides a more truthful and actionable picture than a single average-based metric ever could.

The disciplinary heterogeneity uncovered in our analysis further underscores the necessity of a metric like I3/N. The significant discrepancies in percentile ranks between I3/N and CiteScore, and the subject-specific patterns therein, are highly revealing. The finding that journals in fields like Ecology and Biology tended to be downgraded in I3/N rankings, while those in Humanities and Social Sciences (e.g., Law, Linguistics) were often upgraded, speaks directly to the issue of field normalization. Traditional metrics like JIF and CiteScore often disadvantage fields with lower citation densities and longer citation windows. The use of fractional counting for citations in our I3 calculation, as advocated by Zhou and Leydesdorff (2011), inherently normalizes for differing reference list lengths and citation potentials across disciplines. Consequently, I3/N appears to create a more level playing field, rectifying some of the historical undervaluation of Social Sciences' and Humanities' research within citation-based evaluation systems.

For stakeholders in the scholarly ecosystem, these findings carry profound implications. For researchers and librarians, the Scilit-I3 combination provides a tool for journal selection that is both more comprehensive and less biased. A scholar in a field within Social Sciences, for instance, can now identify journals that, while perhaps having a modest CiteScore, demonstrate strong per-article impact (high I3/N) within the broader Scilit corpus. For publishers, the annual tracking of I3/N distributions across their portfolios (as in Figure 6) offers strategic intelligence. The observed "dilution effect" for publishers undergoing rapid expansion—where the median I3/N decreases as new, lower-impact titles are added—highlights a tangible trade-off between growth and consolidated impact. This data can inform smarter launch and acquisition strategies, moving beyond the simplistic pursuit of raising a journal's JIF. For research managers and policymakers, the I3 framework aligns with the principles of the San Francisco Declaration on Research Assessment (DORA, https://sfdora.org), which advocates for measures that avoid the misuse of journal-based metrics. The ability of I3 to be computed at the article level and then aggregated to the journal, department, or institutional level provides a pathway for responsible evaluation that appreciates the distribution of impact rather than just its mean which is in line of more recent calls for responsible use of metrics, e.g., the Agreement on Reforming Research Assessment (CoARA, https://coara.eu).

However, this approach is not without its limitations. First, the very breadth of Scilit's coverage presents a data quality challenge. The reliance on automated harvesting from diverse sources can lead to metadata inconsistencies, as hinted by the anomalous result for the *International Journal of Systematic and Evolutionary Microbiology*, where incomplete citation data may have inflated its I3 score. While Scilit employs deduplication and disambiguation algorithms, the sheer scale makes perfect data cleanliness an ongoing challenge. Second, the choice of weighting scheme for the I3, while informed by expert advice, remains a value judgment. Assigning a weight of zero to the bottom 50% of papers is a strong statement that entirely disregards the value of publications that may be influential in other ways (e.g., through altmetrics or practice-oriented applications). Alternative weighting schemes could be explored for different evaluative contexts. Third, the cross-disciplinary mapping, though innovative, is an imperfect proxy. Aligning subject categories from WoS, Scopus, and Scilit is a complex task, and residual misalignments could affect the precision of field-normalized comparisons.

Future research should address these limitations and build upon this foundation. A critical next step is to conduct a rigorous, independent audit of Scilit's metadata quality compared to curated databases. Furthermore, exploring the application of the Scilit I3 framework at the meso (institutional) and micro (individual researcher) levels would test its utility for broader research evaluation purposes. Qualitative studies, involving interviews with editors and researchers, could investigate the perceived fairness and usefulness of I3 compared to traditional metrics. Finally, the dynamic response of I3 to emerging trends, such as the rapid growth in AI research noted in our results, warrants longitudinal investigation to understand its sensitivity as an early indicator of shifting scientific paradigms.

In conclusion, this study demonstrates that the integration of the I3 indicator with the Scilit database constitutes a significant step forward in the fields of bibliometrics and scientometrics; the shift may also significantly affect science of science, sociology of science, and science and technology studies (STS). It moves the conversation beyond a fixation on simple averages computed from selective databases toward a more sophisticated, equitable, and responsible model of impact assessment. By embracing a percentile-based, distribution-sensitive metric within an inclusive data infrastructure, the scholarly community can foster a more accurate and diverse representation of scientific progress.

## 6. Conclusion

Through benchmark comparisons of Scilit, I3, and I3/N against JIF and CiteScore, we find that Scilit's broad coverage, combined with I3's distribution-sensitive methodology and normalization, delivers a more granular, equitable, and diagnostically robust framework for assessing journal impact, particularly in interdisciplinary contexts. Scilit significantly expands journal coverage through multi-source aggregation, mitigating the structural biases of traditional databases toward English-language publications and commercially published journals. By integrating complete citation distributions via percentile weighting, I3 and I3/N further reduce the skewness and manipulability inherent in mean-based metrics and allow the decoupling of scale effects from unit-quality assessments, including via journal-quadrant diagnostics. By mapping journals' disciplinary affiliations across datasets, we identify similarities and differences in performance between Scilit-based I3/N and JIF/CiteScore across disciplinary classifications, offering new perspectives for evaluating disciplinary development. Furthermore, I3/N, when combined with fractional counting, corrects for the systematic underestimation of fields with low citation density, such as the Humanities and Social Sciences. Finally, we demonstrate I3/N's ability to capture publisher-level disparities in journal performance, providing decision-making insights for portfolio deployment strategies. Based on these findings, we argue that Scilit and I3/N represent a paradigm shift from "simple averages on selective databases" to "distribution-sensitive evaluation grounded in inclusive data," thereby providing actionable pathways for responsible research assessment. Concurrently, future work should include independent audits of Scilit's metadata quality, extend validation of I3 to institutional and individual levels, and track I3's dynamic sensitivity to emerging fields (e.g., AI) to assess its forward-looking potential.

## 7. Availability of materials and data

All data and code to replicate presented results are publicly deposited at 10.5281/zenodo.18013019.

## 8. Declaration of competing interest

RH is an Associate Editor of Scientometrics. RH and YB are advisory board chairs of the Scilit database. The other authors declare that they are employed by MDPI, the organization that develops and maintains the Scilit database described in this study. The authors declare that this affiliation did not influence the study design, data analysis, interpretation of results, or the decision to publish the manuscript.

## 9. CRediT authorship contribution statement

Haochen Dong: Conceptualization, Data curation, Methodology, Software, Writing – original draft. Qiao Sun: Conceptualization, Data curation, Methodology, Supervision, Writing – original draft. Yanping Mu: Supervision, Writing – original draft. Lu Liao: Supervision, Writing – original draft. Diogo Rodrigues: Writing – original draft. Frank Sauerburger: Writing – original draft. Yi Bu: Supervision, Writing – original draft. Robin Haunschild: Supervision, Writing – original draft.

## 10. Acknowledgments

The authors thank Dr. Shu-Kun Lin and MDPI for technical support, MDPI Author Services for professional English editing, and the Scilit developers (Haijiang Li, Mladen Milutinovic, Darko Todoric, and Fuyong Zhang) for creating the ER diagram. The authors used ChatGPT (OpenAI) as an auxiliary tool for organizing and refactoring parts of the analysis code. All study design, data analysis, interpretation of results, and any remaining errors are the sole responsibility of the authors.